\documentclass[aps,prd,10pt,onecolumn,superscriptaddress]{revtex4-2}

\usepackage[utf8]{inputenc}
\usepackage{amsmath,amssymb}
\usepackage{graphicx}
\usepackage[caption=false]{subfig}
\usepackage{multirow}
\usepackage{appendix}
\usepackage{colortbl}
\usepackage{xcolor}
\usepackage{float}
\usepackage{adjustbox}
\usepackage[colorlinks=true, urlcolor=blue, linkcolor=blue, citecolor=blue]{hyperref}
\providecommand{\Eprint}[2]{\href{#1}{#2}}
\usepackage{rotating}

\begin{document}

\title{Spectroscopic parameters of $B_c$ meson}

\author{Sinem K\"{u}\c{c}\"{u}ky{\i}lmaz}
\email{sinemkucukyilmaz28@gmail.com}
\affiliation{Department of Physics, Faculty of Science,
             Ondokuz Mayis University, 55139, Samsun, T\"{u}rkiye}

\author{Halil Mutuk}
\email{hmutuk@omu.edu.tr}
\affiliation{Department of Physics, Faculty of Science,
             Ondokuz Mayis University, 55139, Samsun, T\"{u}rkiye}


\begin{abstract}

We study the spectroscopy and decays of the $B_c$ meson in a nonrelativistic quark model in which all observables follow from a single set of wave functions, with spin-dependent effects included directly in the bound-state equation. With parameters calibrated to the measured $B_c(1S)$ and $B_c(2S)$ masses, we predict the $S$-, $P$-, and $D$-wave spectrum up to $n=5$, the pseudoscalar and vector decay constants with one-loop QCD corrections, the E1 and M1 radiative widths, and the radial Regge trajectories of ten spin-parity families. The $1P$ multiplet is predicted at 6692--6746~MeV with the normal fine-structure ordering, compatible with lattice QCD and located in the region of the two structures observed by LHCb in the $B_c^+\gamma$ spectrum. The $1D$ states lie near 7008~MeV, below the $BD$ threshold, and all four $D$-wave families are provided. The decay constant $f_{B_c}=507.1$~MeV is consistent with recent QCD sum-rule results and corresponds to $\mathcal{B}(B_c\to\tau\nu_\tau)\approx3\%$. The E1 widths follow the expected angular weights, while their magnitudes are governed mainly by the radial overlap. The $1P\to1S$ (81--112~keV) and $1D\to1P$ (71--93~keV) widths suggest the $1D\to1P\to1S$ cascade as a promising route toward the first $D$-wave $B_c$ state. The Regge trajectories are nearly linear, with slopes decreasing from about 5.8~GeV$^2$ for the $S$-wave to 4.7--4.8~GeV$^2$ for the $D$-wave states. These results may guide forthcoming measurements at the LHC and future $e^+e^-$ facilities.
\end{abstract}

\maketitle

\section{Introduction}
\label{sec:intro}

The $B_c$ meson family, built from a charm quark and a bottom antiquark ($c\bar{b}$), is the only meson system made of two heavy quarks of different flavor. Its states cannot annihilate into gluons or photons, and the levels below the open-flavor $B^{(*)}D^{(*)}$ threshold are not expected to decay strongly. They reach the pseudoscalar ground state mainly through radiative and hadronic transitions, while the ground state itself decays weakly, so the excited levels are likely to be narrow, with comparatively clean decay chains. The unequal constituent masses bring further features that are absent in quarkonium. Charge-conjugation symmetry is lost. As a result, the $^1P_1$ and $^3P_1$ states, and likewise the $^1D_2$ and $^3D_2$ states, may mix into the physical $J^P=1^+$ and $2^-$ states, and the effective charge and magnetic moment entering the E1 and M1 transitions are modified. In addition, the leptonic decay constant $f_{B_c}$ governs the rate of $B_c\to\tau\nu_\tau$. Together with the measured lifetime, this channel constrains new-physics interpretations of the $b\to c\tau\nu$ anomalies~\cite{Alonso:2016oyd,Akeroyd:2017mhr}, and the $B_c$ decay constants also enter the unitarity bounds on $b\to c$ form factors~\cite{CostaiReina:2026deh}. The $B_c$ system thus tests heavy-quark dynamics in a regime complementary to the $c\bar{c}$ and $b\bar{b}$ spectra, with a direct link to flavor physics.

CDF first observed the ground state in $p\bar{p}$ collisions at $\sqrt{s}=1.8$~TeV~\cite{Abe:1998wi}, close to potential-model expectations~\cite{Eichten:1994gt,Gershtein:1994dxw}, and its properties have since been measured precisely at the LHC~\cite{Aaij:2012dd,ParticleDataGroup:2024cfk}. ATLAS reported a first excited state in the $B_c^\pm\pi^+\pi^-$ spectrum consistent with the $2S$ level~\cite{Aad:2014laa}, and CMS and LHCb later resolved the $B_c^+(2S)$ and $B_c^{*+}(2S)$ signals~\cite{CMS:2019uhm,LHCb:2019bem}. ATLAS has since observed the $B_c^{*+}$ directly in $B_c^{*+}\to B_c^+\gamma$, with $M(B_c^{*+})-M(B_c^+)=64.5\pm1.4\,(\text{stat})^{+1.0}_{-1.4}\,(\text{syst})$~MeV~\cite{ATLAS:2026Bcstar}. In the orbital sector, LHCb has observed two peaks in the $B_c^+\gamma$ spectrum at $6704.8\pm5.5\pm2.8$ and $6752.4\pm9.5\pm3.1$~MeV~\cite{LHCb:2025uce}, whose spin-parity assignments are not yet established. To our knowledge, no radiative width of an excited $B_c$ state has been measured, and no $D$-wave state has been reported. The high-luminosity LHC and proposed $e^+e^-$ facilities such as FCC-ee~\cite{FCC:2018byv} and CEPC~\cite{CEPCStudyGroup:2018ghi} are expected to provide much larger samples, and purely leptonic $B_c$ decays may become measurable at LHCb and FCC-ee~\cite{CostaiReina:2026deh}. Predictions that address the spectrum together with its decay observables, including the unexplored $D$-wave sector, may therefore be of use to forthcoming analyses.

The $B_c$ system has been studied in quark models~\cite{Eichten:1994gt,Gershtein:1994dxw,Godfrey:1985xj,Zeng:1994vj,Gupta:1995ps,Ebert:2002pp,Godfrey:2004ya,Fulcher:1998ka,Ebert:2011jc,Monteiro:2016ijw,Soni:2017wvy,Eichten:2019gig,Li:2019tbn,Ortega:2020uvc}, light-front approaches~\cite{Verma:2011yw,Tang:2018myz}, lattice QCD~\cite{Davies:1996gi,Jones:1998ub,Gregory:2009hq,Mathur:2018epb,McNeile:2012qf,Colquhoun:2015oha}, QCD sum rules (QCDSR)~\cite{Colangelo:1992cx,Chabab:1993nz,Kiselev:1993ea,Bagan:1994dy,Wang:2012kw,Wang:2013cdy,Baker:2013mwa,Aliev:2019wcm,Narison:2019tym,Wang:2024fwc,CostaiReina:2026deh,Ozdem:2024qaa}, perturbative and nonrelativistic QCD~\cite{Brambilla:2000db,Penin:2004xi,Onishchenko:2003ui,Lee:2010ts,Chen:2015csa,Tao:2022qxa,Tao:2022hos,Sang:2022tnh,Feng:2022ruy,Tao:2023pzv}, and other approaches~\cite{Ikhdair:2003ry,Ikhdair:2006nx,AbdEl-Hady:1998uiq,Wang:2007av,Wang:2022cxy,Badalian:2007km}. Cornell-type potentials with Breit--Fermi spin-dependent interactions have also been applied recently to $B$ mesons and $bq\bar{q}\bar{q}$ tetraquarks~\cite{Lodha:2026epjp}. The low-lying masses appear broadly consistent across these approaches, presumably because they are anchored to the measured levels, whereas the decay observables show a larger spread. Predictions for $f_{B_c}$ range from about 383 to 564~MeV~\cite{Wang:2012kw,Chaturvedi:2022pmn}. Lattice QCD gives $f_{B_c}=427(6)$~MeV~\cite{McNeile:2012qf} and $f_{B_c^*}/f_{B_c}=0.988(27)$~\cite{Colquhoun:2015oha}, and a recent QCDSR analysis finds $f_{B_c}=499\pm71$~MeV and $f_{B_c^*}=480\pm49$~MeV~\cite{CostaiReina:2026deh}. The E1 width $\Gamma(1\,^3P_0\to1\,^3S_1)$ varies between about 52 and 96~keV~\cite{Li:2019tbn,Bokade:2025lmn,Li:2022bre,Asghar:2019qjl}. Decay constants depend on the wave function at the origin, and radiative widths on the overlap of the initial- and final-state wave functions. Both are sensitive to the short-distance interaction, the spin-dependent forces, relativistic corrections, and the quark masses, and in radiative transitions a 10--20~MeV shift of an intermediate multiplet is further amplified by the $E_\gamma^3$ factor, particularly when the photon energy is small. Decay observables therefore probe the wave functions in ways that the mass spectrum alone does not. Masses and decay observables are therefore best compared when they are obtained from the same wave functions, which is not the case when the spin-dependent interactions are treated as first-order perturbations.

In this work we study the $B_c$ spectrum and its leptonic and radiative decays in a nonrelativistic potential model with the Cornell interaction and the Breit--Fermi spin--spin and spin--orbit interactions, both included nonperturbatively in the radial Schr\"{o}dinger equation. Each state thus has its own radial wave function, which is used for all observables. The parameters are taken from Ref.~\cite{Cakir:2026fzd}, where they were fixed by a fit to the measured $B_c(1S)$ and $B_c(2S)$ masses, and are applied without readjustment, so the orbitally excited levels and the decay observables are predictions. We present (i) the $S$-, $P$-, and $D$-wave spectrum up to $n=5$, including the $^3D_1$, $^3D_2$, $^1D_2$, and $^3D_3$ levels in the $LS$ basis; (ii) the pseudoscalar and vector decay constants with one-loop QCD corrections and $\mathcal{B}(B_c\to\tau\nu_\tau)$; (iii) E1 and M1 widths with overlap integrals evaluated at finite photon momentum and the relativistic recoil factor retained; and (iv) radial Regge trajectories in the $(n,M^2)$ plane for ten spin-parity families.

The model reproduces the $B_c(1S)$ and $B_c(2S)$ masses, with a $2S$--$1S$ excitation energy of 600~MeV close to the measured value of about 597~MeV. The predicted $1P$ multiplet lies at 6692--6746~MeV with the normal fine-structure ordering, and the $1D$ multiplet lies near 7008~MeV, below the $BD$ threshold. We also state the limitations of the framework explicitly. Tensor forces and coupled-channel effects are not included, so the fine-structure splittings are approximate and levels above the $BD$ threshold should be read as bare quark-model states. Moreover, the adopted parameters yield $S$-wave singlet--triplet splittings of only 1--2~MeV, well below the ATLAS value. We regard this measurement as an important input for constraining the spin--spin interaction and indicate in the following sections which results depend on it.

Section~\ref{sec:theory} describes the theoretical framework, Section~\ref{sec:results} presents and discusses the results, and Section~\ref{sec:conclusion} summarizes our conclusions.

\section{Theoretical Framework}
\label{sec:theory}

\subsection{Potential Model}

The $B_c$ meson is composed of two heavy quarks, which makes a nonrelativistic potential-model treatment a suitable starting point for its description. The $c\bar{b}$ system is described by a nonrelativistic Hamiltonian in the center-of-mass frame~\cite{Eichten:1994gt,Soni:2017wvy,Li:2019tbn},
\begin{equation}
  H = M_0 + \frac{\mathbf{p}^{\,2}}{2\mu} + V_{\rm eff}(r),
  \label{eq:Hamiltonian}
\end{equation}
where $M_0=m_b+m_c$ is the quark--antiquark rest-mass sum and
$\mu=m_bm_c/(m_b+m_c)$ is the reduced mass. The meson mass is
$M_{B_c}=M_0+E_{n\ell}$, where $E_{n\ell}$ is the eigenvalue of
$H-M_0$. The effective potential is decomposed as
\begin{equation}
  V_{\rm eff}(r) = V(r) + V_{SS}(r) + V_{LS}(r),
  \label{eq:Veff_def}
\end{equation}
where $V(r)$ is the Cornell interaction, $V_{SS}(r)$ the spin--spin
hyperfine interaction, and $V_{LS}(r)$ the spin--orbit fine-structure
interaction.

The Cornell potential combines the two dominant regimes of
the quark--antiquark interaction~\cite{Eichten:1978tg,Buchmuller:1980su}:
\begin{equation}
  V(r) = -\frac{\kappa\,\alpha_s}{r} + b\,r,
  \label{eq:Cornell}
\end{equation}
where $\kappa=C_F=4/3$ is the color factor of a quark--antiquark pair in the color-singlet channel, $\langle\mathbf{T}_1\cdot\mathbf{T}_2\rangle_{\mathbf{singlet}}=-C_F$; $\alpha_s$ is the strong coupling constant and $b$ is the string tension. At short distances the Coulomb-like term reflects OGE, while at large distances the linear term embodies color confinement. The string tension $b=0.184$~GeV$^2$ of Table~\ref{tab:params} is consistent with lattice determinations of the string tension ($b\approx0.18$--$0.20$~GeV$^2$)~\cite{Bali:2000gf}. 

All three potential contributions are treated nonperturbatively by
direct inclusion in $V_{\rm eff}(r)$. For $V_{SS}(r)$, a perturbative
treatment would force spin-singlet and spin-triplet states to share the
same unperturbed wave function~\cite{Ebert:2002pp,Godfrey:1985xj}.
Nonperturbative inclusion allows singlet and triplet states to acquire
distinct radial wave functions, so that any difference between them is
carried consistently into the decay constants and radiative widths
rather than being excluded by construction. For
$V_{LS}(r)$, direct inclusion in the Schr\"{o}dinger equation ensures
that fine-structure splittings within each $L$-multiplet are generated
self-consistently from the same wave functions used for all other
observables~\cite{Barnes:2005pb,Li:2019tbn}.

The spin--spin (hyperfine) interaction follows from the contact term of the Breit--Fermi Hamiltonian. Using
$\nabla^2(1/r)=-4\pi\delta^{(3)}(\mathbf{r})$ for the vector OGE part
$V_V(r)=-\kappa\alpha_s/r$ of the Cornell
potential~\cite{Lucha:1991vn,Eichten:1994gt},
\begin{equation}
  V_{SS}(r) = \frac{2}{3m_cm_b}\,\nabla^2 V_V(r)\,
  \langle\mathbf{S}_1\cdot\mathbf{S}_2\rangle
  = \frac{8\pi\kappa\alpha_s}{3m_cm_b}\,\delta^{(3)}(\mathbf{r})\,
  \langle\mathbf{S}_1\cdot\mathbf{S}_2\rangle.
  \label{eq:Vss_delta}
\end{equation}
Since a contact interaction cannot be used directly in the
Schr\"odinger equation, the delta function is replaced by a normalized
Gaussian of range $1/\sigma$~\cite{Barnes:2005pb,Soni:2017wvy},
\begin{equation}
  V_{SS}(r) = \frac{8\pi\kappa\alpha_s}{3m_cm_b}
  \left(\frac{\sigma}{\sqrt{\pi}}\right)^3
  e^{-\sigma^2 r^2}\,
  \langle\mathbf{S}_1\cdot\mathbf{S}_2\rangle,
  \label{eq:Vss_gauss}
\end{equation}
which reduces to Eq.~(\ref{eq:Vss_delta}) for $\sigma\to\infty$. The
spin expectation value is
\begin{equation}
  \langle\mathbf{S}_1\cdot\mathbf{S}_2\rangle
  = \frac{1}{2}\bigl[S(S+1) - S_1(S_1+1) - S_2(S_2+1)\bigr]
  = \begin{cases} +\tfrac14, & S=1,\\[2pt] -\tfrac34, & S=0,\end{cases}
  \label{eq:S1S2}
\end{equation}
so that $V_{SS}$ raises the spin-triplet and lowers the spin-singlet
levels. The parameter $\sigma$ sets both the strength and the range of
the hyperfine interaction and thus controls the size of the
singlet--triplet splittings.

For states with $L\geq1$, the spin--orbit interaction lifts the
degeneracy within each $L$-multiplet. For a Lorentz-vector OGE
interaction $V_V(r)=-\kappa\alpha_s/r$ and a Lorentz-scalar confining
interaction $V_S(r)=br$, the nonrelativistic reduction of the
Breit--Fermi Hamiltonian for unequal quark masses
gives~\cite{Eichten:1994gt,Lucha:1991vn,Godfrey:1985xj}
\begin{equation}
  V_{LS}(r) = C_{LS}(r)\,\langle\mathbf{L}\cdot\mathbf{S}\rangle,
  \label{eq:VLS}
\end{equation}
with
\begin{equation}
  C_{LS}(r) = \frac{\kappa\alpha_s}{4r^3}
              \left(\frac{1}{m_c^2}+\frac{1}{m_b^2}+\frac{4}{m_c m_b}\right)
            - \frac{b}{4r}
              \left(\frac{1}{m_c^2}+\frac{1}{m_b^2}\right).
  \label{eq:CLS}
\end{equation}
In the equal-mass limit $m_c=m_b=m$ this reduces to
$C_{LS}(r)=\frac{1}{m^2}\bigl(\frac{2\alpha_s}{r^3}-\frac{b}{2r}\bigr)$,
the form used for charmonium in
Refs.~\cite{Barnes:2005pb,Debastiani:2017msn}.
The first (OGE) term is positive and favors the normal ordering
$M(^3L_{J=L+1})>M(^3L_{J=L-1})$, whereas the second term, arising from
Thomas precession in the scalar confining potential, enters with the
opposite sign and partially cancels it. The two contributions balance at
\begin{equation}
  r^{*} = \left[\frac{\kappa\alpha_s}{b}\,
  \frac{m_c^{-2}+m_b^{-2}+4(m_cm_b)^{-1}}{m_c^{-2}+m_b^{-2}}\right]^{1/2}.
  \label{eq:crossover}
\end{equation}

The tensor interaction is not included in the present analysis. It
vanishes for $S$-wave states and does not shift the spin-weighted
centroid of an $L\geq1$ multiplet, but it is of the same order in
$1/m_q^2$ as the spin--orbit term and contributes to the splittings
within the $^3P_J$ and $^3D_J$ multiplets. The fine-structure
splittings reported below should therefore be regarded as
approximate.

Writing $\Psi_{n\ell m}(\mathbf{r}) = r^{-1}\phi_{n\ell}(r)\,Y_\ell^m(\hat{\mathbf{r}})$, the Schr\"{o}dinger equation reduces to the radial
eigenvalue problem
\begin{equation}
  \left[
    -\frac{1}{2\mu}\frac{d^2}{dr^2}
    + \frac{\ell(\ell+1)}{2\mu r^2}
    + V_{\rm eff}(r)
  \right]\phi_{n\ell}(r)
  = E_{n\ell}\,\phi_{n\ell}(r),
  \label{eq:radial_SE}
\end{equation}
solved numerically using the Numerov algorithm on a uniform radial
grid, subject to the regularity condition $\phi_{n\ell}(0)=0$ at
the origin and $\phi_{n\ell}(r_0)=0$ at an outer cutoff $r_0$ chosen
large enough that the bound-state wave function is negligible at the
boundary. The eigenvalues are stable to within 0.1~MeV for
$r_0\gtrsim30$~GeV$^{-1}$. The eigenvalue $E_{n\ell}$ thus incorporates the Cornell interaction together with the spin--spin and spin--orbit splittings, and the meson mass follows from $M_{B_c}=m_b+m_c+E_{n\ell}$ as defined below Eq.~(\ref{eq:Hamiltonian}).

\subsection{Decay Constants}
\label{subsec:decay_const}

The leptonic decay constants of the pseudoscalar ($B_c$) and vector
($B_c^*$) states are defined through the hadronic matrix elements
\begin{align}
  \langle 0 | \bar{b}\,\gamma^\mu\gamma_5\,c | P(k) \rangle
  &= i f_P k^\mu,
  \label{eq:fP_def} \\
  \langle 0 | \bar{b}\,\gamma^\mu\,c | V(k,\varepsilon) \rangle
  &= f_V M_V \varepsilon^\mu,
  \label{eq:fV_def}
\end{align}
where $f_P$ ($f_V$) is the pseudoscalar (vector) decay constant, and
$k^\mu$, $\varepsilon^\mu$, $M_V$ have their standard meanings. These
constants govern the weak annihilation amplitudes and enter the leptonic
widths $\Gamma(B_c^+\to\ell^+\nu_\ell)\propto f_{B_c}^2|V_{cb}|^2m_\ell^2
M_{B_c}$ and $\Gamma(B_c^{*+}\to\ell^+\nu_\ell)\propto f_{B_c^*}^2
|V_{cb}|^2M_{B_c^*}^3$.

In the nonrelativistic limit, $f_{P/V}$ is evaluated via the Van
Royen--Weisskopf formula supplemented by the one-loop QCD radiative
correction~\cite{VanRoyen:1967nq,Braaten:1995ej,Soni:2017wvy,Li:2019tbn,
Gershtein:1994jw}:
\begin{equation}
  f_{P/V}^{\,2}
  = \frac{3\,|R_{nS}(0)|^2}{\pi\,M_{nS}}\,\bar{C}^{\,2}(\alpha_s),
  \label{eq:fPV}
\end{equation}
where $R_{nS}(0)=\lim_{r\to0}\phi_{nS}(r)/r$ is the radial wave function
at the origin and $M_{nS}$ the meson mass. Since $R_{n\ell}(0)=0$ for
$\ell\geq1$, this formula applies only to $S$-wave states. The
dependence on $|R_{nS}(0)|^2$ reflects the fact that decay constants
probe the wave function at zero separation, a purely short-distance
quantity controlled by the Coulombic region of the interaction; they
are consequently more sensitive to $\alpha_s$ and the constituent quark
masses than to the string tension $b$ that dominates the mass
eigenvalues.

The one-loop QCD correction factor $\bar{C}(\alpha_s)$
is~\cite{Braaten:1995ej,Soni:2017wvy,Gershtein:1994jw}:
\begin{equation}
  \bar{C}(\alpha_s)
  = 1 - \frac{\alpha_s}{\pi}
    \left[
      \delta^{(P,V)} - \frac{m_b - m_c}{m_b + m_c}\,\ln\!\left(\frac{m_b}{m_c}\right)
    \right],
  \label{eq:QCDcorr}
\end{equation}
with $\delta^{(P)}=2$ for pseudoscalar and $\delta^{(V)}=8/3$ for
vector states~\cite{Braaten:1995ej}. The logarithmic term
$\frac{m_b-m_c}{m_b+m_c}\ln(m_b/m_c)$ is specific to the unequal-mass
$B_c$ system (it vanishes identically for $c\bar{c}$ and $b\bar{b}$)
and arises from the one-loop vertex correction with unequal quark
masses; it reduces the effective radiative correction relative to a
symmetric quarkonium system. Because $\delta^{(V)}>\delta^{(P)}$, the
radiative correction suppresses the vector decay constant more strongly
than the pseudoscalar, so that at one-loop order it reduces the ratio
$f_{B_c^*}/f_{B_c}$ below unity.

\subsection{Radiative Transitions}
\label{subsec:radiative}

Electromagnetic transitions between $B_c$ states are computed at
leading order in the multipole expansion as electric-dipole (E1) and
magnetic-dipole (M1) transitions, subject to the selection rules
\begin{align}
  \text{E1:} & \quad \Delta S = 0,\quad \Delta L = \pm 1, \label{eq:E1_selection}\\
  \text{M1:} & \quad \Delta S = \pm 1,\quad \Delta L = 0. \label{eq:M1_selection}
\end{align}
Higher-order multipole contributions ($E2$, $M2$, etc.) are suppressed
by additional powers of $E_\gamma r$ and are neglected throughout.

\subsubsection{E1 transitions}
The partial width for $n_i^{\,2S+1}L_{J_i}\to\gamma+n_f^{\,2S+1}L_{J_f}$
is~\cite{Soni:2017wvy,Li:2019tbn}
\begin{equation}
  \Gamma_{E1}
  = \frac{4\alpha\langle e_Q\rangle^2}{3}
    \,E_{\gamma}^{3}
    \,\frac{E_f}{M_i}
    \,C_{fi}
    \,|\epsilon_{fi}|^{2},
  \label{eq:E1_width}
\end{equation}
where $\alpha$ is the fine-structure constant, $E_\gamma=(M_i^2-M_f^2)
/(2M_i)$ the photon energy, $E_f/M_i$ the relativistic recoil factor,
$C_{fi}$ the angular statistical factor, and $\epsilon_{fi}$ the radial
overlap integral. The spin-conservation constraint $\Delta S=0$ is
enforced by the choice of initial and final states. The effective
charge, weighted by the reduced-mass kinematics of the unequal-mass
$c\bar{b}$ system~\cite{Godfrey:1985xj,Ebert:2002pp}, is
\begin{equation}
  \langle e_Q \rangle
  = \frac{m_b\,e_c + m_c\,e_b}{m_b + m_c},
  \label{eq:eff_charge}
\end{equation}
with $e_b=-1/3$ and $e_c=+2/3$; numerically $\langle e_Q\rangle
\approx0.416$ for the present parameter set. The angular factor is
\begin{equation}
  C_{fi}
  = \max(L_i,L_f)\,(2J_f+1)
    \begin{Bmatrix}
      L_f & J_f & S \\
      J_i & L_i & 1
    \end{Bmatrix}^{2}.
  \label{eq:Cfi}
\end{equation}

The radial overlap integral is evaluated retaining the full
recoil-corrected structure rather than truncating at the
long-wavelength limit. For the photon energies encountered in the
$B_c$ spectrum ($E_\gamma\sim100$--$1000$~MeV) and a typical meson
size $\langle r\rangle\sim0.3$--$0.5$~fm, the product $E_\gamma r$
is not small, so the standard long-wavelength approximation
$\epsilon_{fi}\approx\langle r\rangle$ is not justified. The
full expression is~\cite{Soni:2017wvy,Li:2019tbn}
\begin{equation}
  \epsilon_{fi}
  = \frac{3}{E_{\gamma}}
    \int_0^{\infty} dr\,\phi_{n_i L_i}(r)\,\phi_{n_f L_f}(r)
    \left[
      \frac{E_{\gamma}\,r}{2}\,j_0\!\left(\frac{E_{\gamma}\,r}{2}\right)
      - j_1\!\left(\frac{E_{\gamma}\,r}{2}\right)
    \right],
  \label{eq:overlap_E1}
\end{equation}
with $j_0(x)=\sin(x)/x$ and $j_1(x)=\sin(x)/x^2-\cos(x)/x$. In the
limit $E_\gamma r\to0$ one recovers the long-wavelength result
$\epsilon_{fi}\to\int_0^\infty dr\,\phi_{n_iL_i}(r)\,r\,\phi_{n_fL_f}(r)$.

\subsubsection{M1 transitions}
The M1 partial width for $n_i^{\,2S_i+1}L_{J_i}\to\gamma+n_f^{\,2S_f+1}
L_{J_f}$ is~\cite{Soni:2017wvy,Li:2019tbn}
\begin{equation}
  \Gamma_{M1}
  = \frac{4\alpha\,\mu_q^2}{3}
    \,\frac{2J_f+1}{2L+1}
    \,E_{\gamma}^{3}
    \,\frac{E_f}{M_i}
    \,|m_{fi}|^{2},
  \label{eq:M1_width}
\end{equation}
where $L=L_i=L_f$ (enforced by the M1 selection rule $\Delta L=0$).
The effective magnetic dipole moment is~\cite{Godfrey:1985xj,Ebert:2002pp,
Soni:2017wvy,Li:2019tbn}
\begin{equation}
  \mu_q
  = \frac{m_b\,e_c + m_c\,e_b}{2\,m_b\,m_c},
  \label{eq:mu_eff}
\end{equation}
and the M1 radial overlap integral is
\begin{equation}
  m_{fi}
  = \int_0^{\infty} dr\,\phi_{n_f L}(r)\,\phi_{n_i L}(r)\,
    j_0\!\left(\frac{E_{\gamma}\,r}{2}\right).
  \label{eq:overlap_M1}
\end{equation}
For allowed transitions ($n_i=n_f$), $m_{fi}$ is nearly maximal in the $E_\gamma\to0$ limit and the width is controlled almost entirely by $E_\gamma^3$. For hindered transitions ($n_i\neq n_f$), singlet and triplet states with different radial quantum numbers would be orthogonal in the limit $V_{SS}=0$ and $E_\gamma\to0$; both the Gaussian hyperfine interaction, which distorts singlet and triplet wave functions differently, and the finite photon momentum entering
through $j_0(E_\gamma r/2)$ break this orthogonality. The relative weight of these two effects depends on the strength of the hyperfine interaction and is discussed in Sec.~\ref{subsec:rad_results}.

\subsection{Regge Trajectories}
\label{subsec:regge_theory}

Linear confinement leads to radial level spacings that are
approximately uniform in $M^2$, although for heavy quarks the Coulombic
interaction produces deviations at low $n$. Successive excitations within a given spin-parity
family therefore trace nearly linear trajectories in the $(n,M^2)$
plane. The trajectory is parametrized as
\begin{equation}
  M^2 = \beta_0 + \beta\,n,
  \label{eq:Regge}
\end{equation}
where $n=1,2,3,4,5$ is the radial quantum number ($n=1$ for the
ground state), $\beta$ is the slope in GeV$^2$, and $\beta_0$ the
intercept~\cite{Ebert:2011jc,Soni:2017wvy,Li:2019tbn}. Separate
trajectories are constructed for the $n\,^1S_0$ and $n\,^3S_1$
families, for $n\,^1P_1$, $n\,^3P_0$, $n\,^3P_1$, and
$n\,^3P_2$, and for $n\,^1D_2$, $n\,^3D_1$, $n\,^3D_2$, and
$n\,^3D_3$. For each family, the squared masses from the numerical solution
of Eq.~(\ref{eq:radial_SE}) are fitted by linear least squares
to extract $\beta$, $\beta_0$, and the coefficient of determination
$R^2$. The fitted parameters provide a compact characterization of
the radial spectrum and a basis for comparison with other approaches.

\section{Results and Discussion}
\label{sec:results}

\subsection{Mass Spectrum}
\label{subsec:massspectrum}

The masses of the $S$-, $P$-, and $D$-wave states are listed in Tables~\ref{tab:swave}--\ref{tab:dwave_all}, together with the experimental data and results from quark models, lattice QCD, the Bethe--Salpeter approach, and QCDSR. The parameters of Table~\ref{tab:params} were determined in Ref.~\cite{Cakir:2026fzd} from a fit to the measured $B_c(1S)$ and $B_c(2S)$ masses. These two levels therefore check the calibration, and all other states are predictions. Our $1S$ and $2S$ masses differ by about 1~MeV from those of Ref.~\cite{Cakir:2026fzd} because of differences in the numerical implementation. Levels above the $BD$ threshold ($\simeq7144$~MeV) are bare quark-model states.

\begin{table}[!h]
\centering
\caption {Quark-model parameters adopted in the present work, taken from Ref.~\cite{Cakir:2026fzd}.} 
\label{tab:params}
\setlength{\tabcolsep}{8pt}
\begin{tabular}{cccccc}
\hline\hline
$m_b$ & $m_c$ & $\alpha_s$ & $b$ & $\sigma$ \\
(GeV) & (GeV) &            & (GeV$^2$) & (GeV)  \\
\hline
4.605 & 1.537 & 0.4268 & 0.1840 & 0.2614 \\
\hline\hline
\end{tabular}
\end{table}

\subsubsection{$S$-wave}

The ground-state mass $M(1\,^1S_0)=6270$~MeV lies about 4~MeV below the PDG value~\cite{ParticleDataGroup:2024cfk} and is compatible with the lattice QCD result $6276\pm7$~MeV~\cite{Mathur:2018epb}. The $2\,^1S_0$ mass of 6870~MeV agrees with the CMS and LHCb measurements~\cite{CMS:2019uhm,LHCb:2019bem} within about 1~MeV. The resulting $2S$--$1S$ excitation energy of 600~MeV, close to the measured value of about 597~MeV, is governed mainly by the Coulombic and linear terms of the Cornell potential. Most approaches in Table~\ref{tab:swave} describe the $2\,^1S_0$ level comparably, whereas Ref.~\cite{Asghar:2019qjl} yields an excitation energy of about 423~MeV.

For $n\geq3$ our masses lie above most other predictions, by about 20--60~MeV at $n=3$ and by up to about 200~MeV at $n=5$. The results of Ref.~\cite{Ebert:2011jc} approach ours with increasing $n$ and coincide within 1~MeV for the $5\,^1S_0$ state. This spread reflects the different treatments of confinement, to which highly excited levels are most sensitive.

The singlet--triplet splittings $\Delta M(nS)=M(n\,^3S_1)-M(n\,^1S_0)$ are only 1--2~MeV, well below the 64.5~MeV measured by ATLAS~\cite{ATLAS:2026Bcstar}. This reflects the small hyperfine strength implied by $\sigma$, which was not constrained by spin-triplet data. Combined with the PDG $B_c$ mass, the ATLAS result gives $M(B_c^*)\simeq6339$~MeV, close to the lattice QCD value $6331\pm7$~MeV~\cite{Mathur:2018epb} and to the upper end of the 6322--6338~MeV range of most models in Table~\ref{tab:swave}, whereas our $1\,^3S_1$ mass is 6272~MeV. A consistent treatment would require a joint readjustment of $\sigma$ and the other parameters, which we leave for future work. The $S$-wave triplet masses are therefore the least constrained part of the spectrum.

\begin{table}[h]
\centering
\caption{$S$-wave mass spectrum of the $B_c$ meson (MeV). $^{a}$Obtained from the PDG $B_c$ mass~\cite{ParticleDataGroup:2024cfk} and the $B_c^*$--$B_c$ splitting measured by ATLAS~\cite{ATLAS:2026Bcstar}.}
\footnotesize
\setlength{\tabcolsep}{3pt}
\renewcommand{\arraystretch}{1.1}
\begin{tabular}{c c c c c c c c c c c c}
\hline\hline\\[0.1pt]
$n\,^{2S+1}L_J$ & This Work & \shortstack{Experiment~\cite{ParticleDataGroup:2024cfk}} & \shortstack{\cite{Bokade:2025lmn}} & \shortstack{\cite{Li:2022bre}} & \shortstack{\cite{Li:2019tbn}} & \shortstack{\cite{Li:2023wgq}} & \shortstack{\cite{Asghar:2019qjl}} & \shortstack{\cite{Ebert:2011jc}} & \shortstack{\cite{Mathur:2018epb}} & \shortstack{\cite{Wang:2022cxy}} & \shortstack{\cite{Wang:2012kw}} \\[4pt]
\hline\\[1pt]
$1\,^1S_0$ & 6270 & 6274.47 & 6275.8 & 6269 & 6271 & 6271 & 6318 & 6272 & $6276 \pm 7$ & 6277 & --- \\
$2\,^1S_0$ & 6870 & $6871.2 \pm 1.0$ & 6869.8 & 6886 & 6871 & 6855 & 6741 & 6842 & --- & 6867 & --- \\
$3\,^1S_0$ & 7280 & --- & 7252.1 & 7261 & 7239 & 7220 & 7014 & 7226 & --- & 7228 & --- \\
$4\,^1S_0$ & 7622 & --- & 7545.1 & 7551 & 7540 & 7496 & 7239 & 7585 & --- & --- & --- \\
$5\,^1S_0$ & 7927 & --- & 7785.4 & 7790 & 7805 & 7722 & --- & 7928 & --- & --- & --- \\[4pt]
$1\,^3S_1$ & 6272 & $6339\pm2^{\,a}$ & 6336.1 & 6322 & 6326 & 6338 & 6336 & 6333 & $6331 \pm 7$ & 6332 & $6331 \pm 47$ \\
$2\,^3S_1$ & 6871 & --- & 6905.6 & 6907 & 6890 & 6886 & 6747 & 6882 & --- & 6911 & --- \\
$3\,^3S_1$ & 7280 & --- & 7278.4 & 7275 & 7252 & 7240 & 7018 & 7258 & --- & 7272 & --- \\
$4\,^3S_1$ & 7623 & --- & 7565.9 & 7561 & 7550 & 7512 & 7242 & 7609 & --- & --- & --- \\
$5\,^3S_1$ & 7928 & --- & 7802.6 & 7798 & 7813 & 7735 & --- & 7947 & --- & --- & --- \\[4pt]
\hline\hline
\end{tabular}
\label{tab:swave}
\end{table}

\subsubsection{$P$-wave}

The $1P$ multiplet lies at 6692--6746~MeV, in line with relativistic quark models~\cite{Ebert:2011jc}, the Salpeter equation~\cite{Wang:2022cxy}, QCDSR~\cite{Wang:2012kw}, and modified Godfrey--Isgur schemes~\cite{Li:2023wgq}. The $1\,^3P_0$ and $1\,^1P_1$ masses are compatible within one standard deviation with the lattice QCD results $6712\pm25$ and $6736\pm24$~MeV of Ref.~\cite{Mathur:2018epb}.

The normal ordering $M(^3P_0)<M(^3P_1)<M(^3P_2)$ holds at every radial excitation. The splittings $M(^3P_1)-M(^3P_0)\simeq17$--18~MeV and $M(^3P_2)-M(^3P_1)\simeq33$--36~MeV follow the $1:2$ ratio expected from $\langle\mathbf{L}\cdot\mathbf{S}\rangle=-2,-1,+1$, and each $n\,^1P_1$ mass lies within 2~MeV of the spin-weighted centroid of the $n\,^3P_J$ triplet. Both relations hold at first order in the spin--orbit interaction when the $P$-wave hyperfine term is negligible. Their approximate validity indicates that the higher-order effects of the nonperturbative treatment are small for the $P$-wave states.

The total splitting $M(1\,^3P_2)-M(1\,^3P_0)=54$~MeV lies below the 57--73~MeV range of Refs.~\cite{Ebert:2011jc,Li:2019tbn,Wang:2022cxy,Bokade:2025lmn,Li:2022bre,Li:2023wgq} and decreases slowly to 50~MeV for the $5P$ states. According to Eq.~(\ref{eq:crossover}), the crossover radius $r^*\approx0.5$~fm is comparable to the mean interquark separation of the $1P$ states. The OGE and Thomas-precession contributions therefore partially cancel and leave a positive but moderate spin--orbit coupling. At higher radial excitation the wave function extends further into the confining region, where the Thomas-precession term gains weight. The $1P$ fine structure is thus linked to $\kappa\alpha_s/b$, and a measurement of two or more members of the multiplet would probe this ratio directly.

The two LHCb peaks at $6704.8\pm5.5\pm2.8$ and $6752.4\pm9.5\pm3.1$~MeV~\cite{LHCb:2025uce} lie in the region of the $1P$ multiplet. Only the $J^P=1^+$ states decay directly to $B_c\gamma$, whereas transitions to $B_c^*\gamma$ appear shifted down by the $1S$ hyperfine splitting, about 64.5~MeV~\cite{ATLAS:2026Bcstar}, when the soft photon is not reconstructed. If both peaks are attributed to the $1^+$ states, the average of our $1\,^3P_1$ and $1\,^1P_1$ masses, 6718~MeV, lies within about 11~MeV of the average of the peaks, 6729~MeV. The predicted separation of 16~MeV is smaller than the observed one. Since $^1P_1$--$^3P_1$ mixing tends to increase this separation, our $LS$-basis masses may serve as input for mixing analyses, while a definite assignment will likely require spin-parity measurements. The $2\,^1P_1$ mass agrees with Refs.~\cite{Ebert:2011jc,Soni:2017wvy,Li:2019tbn,Wang:2022cxy,Bokade:2025lmn,Li:2022bre,Li:2023wgq} within about 20~MeV, and up to $n=4$ the $P$-wave levels remain within about 50~MeV of the relativistic results of Ref.~\cite{Ebert:2011jc}.

\begin{table*}[htbp]
\centering
\caption{$P$-wave mass spectrum of the $B_c$ meson (MeV). LHCb has reported two peaks in the $B_c^+\gamma$ spectrum at $6704.8\pm5.5\pm2.8$ and $6752.4\pm9.5\pm3.1$~MeV~\cite{LHCb:2025uce}, whose spin-parity assignments are not established (see text).}
\label{tab:pwave_all}
\renewcommand{\arraystretch}{1.1}
\begin{small}
\begin{tabular}{cccccccccccc}
\hline\hline
\rule{0pt}{14pt}$n^{2S+1}L_J$ & This Work & \cite{Bokade:2025lmn} & \cite{Li:2022bre} & \cite{Li:2019tbn} & \cite{Li:2023wgq} & \cite{Asghar:2019qjl} & \cite{Ebert:2011jc} & \cite{Soni:2017wvy} & \shortstack{\cite{Mathur:2018epb}} & \shortstack{\cite{Wang:2022cxy}} & \shortstack{\cite{Wang:2012kw}} \\[4pt] \hline
\rule{0pt}{12pt}$1\,^3P_0$ & 6692 & 6703.9 & 6712 & 6714 & 6701 & 6631 & 6699 & 6686 & $6712 \pm 25$ & 6705 & --- \\
$2\,^3P_0$ & 7119 & 7111.4 & 7118 & 7107 & 7097 & 6915 & 7094 & 7146 & --- & 7112 & --- \\
$3\,^3P_0$ & 7472 & 7423.2 & 7427 & 7420 & 7393 & 7147 & 7474 & 7536 & --- & 7408 & --- \\
$4\,^3P_0$ & 7784 & 7678.9 & 7682 & 7693 & 7633 & 7350 & 7817 & 7885 & --- & --- & --- \\
$5\,^3P_0$ & 8069 & 7896.4 & 7899 & --- & --- & --- & --- & 8207 & --- & --- & --- \\[4pt]
$1\,^3P_1$ & 6710 & --- & --- & --- & --- & --- & 6750 & 6712 & --- & 6748 & --- \\
$2\,^3P_1$ & 7136 & --- & --- & --- & --- & --- & 7134 & 7173 & --- & 7149 & --- \\
$3\,^3P_1$ & 7489 & --- & --- & --- & --- & --- & 7510 & 7565 & --- & 7442 & --- \\
$4\,^3P_1$ & 7801 & --- & --- & --- & --- & --- & 7853 & 7915 & --- & --- & --- \\
$5\,^3P_1$ & 8086 & --- & --- & --- & --- & --- & --- & 8237 & --- & --- & --- \\[4pt]
$1\,^1P_1$ & 6726 & 6751.9 & 6761 & 6776 & 6754 & 6650 & 6750 & 6706 & $6736 \pm 24$ & 6739 & $6737 \pm 56$ \\
$2\,^1P_1$ & 7152 & 7157.9 & 7156 & 7150 & 7133 & 6930 & 7147 & 7168 & --- & 7144 & --- \\
$3\,^1P_1$ & 7505 & 7467.7 & 7458 & 7458 & 7421 & 7162 & 7510 & 7559 & --- & 7440 & --- \\
$4\,^1P_1$ & 7817 & 7721.4 & 7708 & 7727 & 7656 & 7364 & 7853 & 7908 & --- & --- & --- \\
$5\,^1P_1$ & 8102 & 7936.6 & 7921 & --- & --- & --- & --- & 8230 & --- & --- & --- \\[4pt]
$1\,^3P_2$ & 6746 & 6772.6 & 6783 & 6787 & 6773 & 6665 & 6761 & 6712 & --- & 6762 & --- \\
$2\,^3P_2$ & 7171 & 7178.4 & 7175 & 7160 & 7148 & 6946 & 7157 & 7173 & --- & 7163 & --- \\
$3\,^3P_2$ & 7523 & 7487.2 & 7476 & 7464 & 7434 & 7176 & 7524 & 7565 & --- & 7456 & --- \\
$4\,^3P_2$ & 7834 & 7739.6 & 7724 & 7732 & 7667 & 7379 & 7867 & 7915 & --- & --- & --- \\
$5\,^3P_2$ & 8119 & 7953.7 & 7936 & --- & --- & --- & --- & 8237 & --- & --- & --- \\
\hline\hline
\end{tabular}
\end{small}
\end{table*}

\subsubsection{$D$-wave}

No $D$-wave $B_c$ state has been observed, so the results in Table~\ref{tab:dwave_all} are predictions. The four $1D$ levels lie at 7008--7009~MeV, close to the results of Ref.~\cite{Soni:2017wvy} and within about 45~MeV of most other references. Lying below the $BD$ threshold, they are expected to decay mainly through the radiative transitions discussed in Sec.~\ref{subsec:rad_results}.

The near degeneracy of the $1D$ levels suggests that the OGE and Thomas-precession contributions nearly cancel at the separations sampled by the $1D$ wave functions. For $n\geq2$ a small normal splitting develops, with $M(^3D_3)-M(^3D_1)$ between 5 and 11~MeV. Since it results from a near cancellation and the tensor interaction is not included, this splitting should be regarded as approximate. Refs.~\cite{Ebert:2011jc,Li:2019tbn,Wang:2022cxy,Bokade:2025lmn,Li:2022bre,Li:2023wgq} also find small splittings of 3--21~MeV for the $1D$ multiplet, whereas Ref.~\cite{Soni:2017wvy} obtains an inverted ordering. The $D$-wave fine structure thus appears to be a sensitive probe of the balance between the vector and scalar components of the interaction.

We provide all four $^3D_1$, $^3D_2$, $^1D_2$, and $^3D_3$ levels for $n=1$--5. The $^3D_2$ levels are absent from the comparison references of Table~\ref{tab:dwave_all}, and the separate $LS$-basis $^1D_2$ and $^3D_2$ masses provide input for the physical $J^P=2^-$ states that arise from their mixing.

\begin{table*}[!h]
\centering
\caption{$D$-wave mass spectrum of the $B_c$ meson (MeV).}
\label{tab:dwave_all}
\setlength{\tabcolsep}{5pt}
\renewcommand{\arraystretch}{1.1}
\begin{small}
\begin{tabular}{cccccccccc}
\hline
\rule{0pt}{14pt}$n^{2S+1}L_J$ & This Work &\cite{Bokade:2025lmn} &\cite{Li:2022bre} & \cite{Li:2019tbn} & \cite{Li:2023wgq} & \cite{Asghar:2019qjl} & \cite{Ebert:2011jc} & \cite{Soni:2017wvy}& \shortstack{\cite{Wang:2022cxy}} \\[4pt] \hline

\rule{0pt}{12pt}$1\,^3D_1$ & 7008 & 7038.3 & 7037 & 7020 & 7023 & 6841 & 7021 & 6998 & 7014 \\
$2\,^3D_1$ & 7371 & 7361.8 & 7357 & 7336 & 7327 & 7080 & 7392 & 7403 & 7335 \\
$3\,^3D_1$ & 7691 & 7625.7 & 7619 & 7611 & 7573 & 7289 & 7732 & 7762 & --- \\
$4\,^3D_1$ & 7981 & 7849.3 & 7842 & --- & --- & 7478 & --- & 8091 & --- \\
$5\,^3D_1$ & 8251 & 8043.3 & --- & --- & --- & --- & --- & --- & --- \\[4pt]
$1\,^3D_2$ & 7009 & --- & --- & --- & --- & --- & --- & --- & --- \\
$2\,^3D_2$ & 7373 & --- & --- & --- & --- & --- & --- & --- & --- \\
$3\,^3D_2$ & 7694 & --- & --- & --- & --- & --- & --- & --- & --- \\
$4\,^3D_2$ & 7985 & --- & --- & --- & --- & --- & --- & --- & --- \\
$5\,^3D_2$ & 8256 & --- & --- & --- & --- & --- & --- & --- & --- \\[4pt]

$1\,^1D_2$ & 7008 & 7053.0 & 7046 & 7032 & 7039 & 6845 & 7026 & 6994 & 7025 \\
$2\,^1D_2$ & 7374 & 7375.9 & 7365 & 7347 & 7340 & 7084 & 7400 & 7401 & --- \\
$3\,^1D_2$ & 7695 & 7639.2 & 7627 & 7623 & 7584 & 7293 & 7743 & 7762 & --- \\
$4\,^1D_2$ & 7987 & 7862.1 & 7849 & --- & --- & 7482 & --- & 8093 & --- \\
$5\,^1D_2$ & 8258 & 8055.4 & --- & --- & --- & --- & --- & --- & --- \\[4pt]

$1\,^3D_3$ & 7009 & 7041.4 & 7042 & 7030 & 7042 & 6847 & 7029 & 6990 & 7035 \\
$2\,^3D_3$ & 7376 & 7371.4 & 7364 & 7348 & 7344 & 7087 & 7405 & 7399 & --- \\
$3\,^3D_3$ & 7698 & 7638.8 & 7627 & 7625 & 7589 & 7296 & 7750 & 7761 & --- \\
$4\,^3D_3$ & 7991 & 7864.4 & 7850 & --- & --- & 7489 & --- & 8092 & --- \\
$5\,^3D_3$ & 8262 & 8059.4 & --- & --- & --- & --- & --- & --- & ---\\
\hline
\end{tabular}
\end{small}
\end{table*}

\subsection{Decay Constants}
\label{subsec:dc_results}

Table~\ref{tab:decay_constant} lists the decay constants obtained from Eq.~(\ref{eq:fPV}). The ground-state value $f_{B_c}=507.1$~MeV lies within the 433--564~MeV range of the model calculations~\cite{Soni:2017wvy,Chaturvedi:2022pmn,Bokade:2025lmn,2209.06724,2003.08491,Patel:2008mv}. It is consistent, within uncertainties, with the estimate $460\pm60$~MeV of Ref.~\cite{Gershtein:1994jw} and with the value $499\pm71$~MeV of a recent QCDSR analysis~\cite{CostaiReina:2026deh}. The lattice QCD result $f_{B_c}=427(6)$~MeV~\cite{McNeile:2012qf} is about 16\% lower. Many potential-model results based on the Van Royen--Weisskopf relation show a similar excess, likely related to the sensitivity of $|R_{nS}(0)|^2$ to the short-distance interaction and to sizable higher-order QCD corrections~\cite{Onishchenko:2003ui,Chen:2015csa,Feng:2022ruy,Tao:2023pzv}. With $\tau_{B_c}=0.510$~ps and $|V_{cb}|=0.041$, our value gives $\mathcal{B}(B_c\to\tau\nu_\tau)\approx3\%$.

For the vector state we obtain $f_{B_c^*}=449.5$~MeV, and the ratio $f_{B_c^*}/f_{B_c}\approx0.89$ is nearly independent of $n$. Because the hyperfine interaction is weak in the present parametrization, the singlet and triplet wave functions at the origin almost coincide. The ratio is then fixed mainly by the one-loop factor $\bar{C}_V/\bar{C}_P\simeq0.887$ of Eq.~(\ref{eq:QCDcorr}). It lies below the HPQCD result $0.988(27)$~\cite{Colquhoun:2015oha} and should be regarded as an estimate of the one-loop effect alone.

The decay constants decrease monotonically with $n$, by about 28\% from $1S$ to $2S$ and by about 5\% from $4S$ to $5S$, as the radial wave function spreads away from the origin. All models of Table~\ref{tab:decay_constant} that provide excited states show the same trend.

\begin{table*}[htbp]
\centering
\caption{Pseudoscalar $f_P$ and vector $f_V$ decay constants of the
$B_c$ meson (MeV).}
\label{tab:decay_constant}
\setlength{\tabcolsep}{5pt}
\renewcommand{\arraystretch}{1.1}
\begin{small}
\begin{tabular}{cccccccccc}
\hline\hline
\rule{0pt}{14pt}$n^{2S+1}L_J$ & This Work & \cite{Bokade:2025lmn} & \cite{2209.06724} & \cite{Chaturvedi:2022pmn} & \cite{2003.08491} & \cite{Soni:2017wvy} & \cite{Patel:2008mv} & \cite{Gershtein:1994jw} & \begin{tabular}{c}\cite{Wang:2012kw}\end{tabular} \\[4pt] \hline
\rule{0pt}{12pt}$1\,^1S_0$ & 507.1 & 530.3 & 439 & 564.1 & 484.4 & 433.0 & 465 & $460 \pm 60$ & $383 \pm 27$ \\
$2\,^1S_0$ & 363.1 & 434   & 282 & 451.6 & 347.2 & 355.5 & 361 & --- & --- \\
$3\,^1S_0$ & 319.7 & 388.7 & 237 & 410.9 & 306.4 & 325.7 & 319 & --- & --- \\
$4\,^1S_0$ & 295.9 & 358.7 & --- & 386.2 & 284.0 & 307.5 & 293 & --- & --- \\
$5\,^1S_0$ & 279.8 & 335.5 & --- & 368.7 & 268.8 & 294.4 & 275 & --- & --- \\[4pt]
$1\,^3S_1$ & 449.5  & 513.4 & 417 & 531.9 & 404.9 & 434.6 & 435 & $460\pm60$ & $384\pm32$ \\
$2\,^3S_1$ & 321.8  & 421.1 & 297 & 424.5 & 303.8 & 356.4 & 337 & ---        & ---        \\
$3\,^3S_1$ & 283.5  & 377.4 & 257 & 385.5 & 270.3 & 326.4 & 297 & ---        & ---        \\
$4\,^3S_1$ & 262.4  & 348.5 & --- & 362.3 & 251.5 & 308.1 & 273 & ---        & ---        \\
$5\,^3S_1$ & 248.1  & 326   & --- & 345.6 & 238.7 & 295.0 & 256 & ---        & ---        \\
\hline\hline
\end{tabular}
\end{small}
\end{table*}

\subsection{Radiative Transitions}
\label{subsec:rad_results}

Tables~\ref{tab:sw_e1}--\ref{tab:m1_sw} list the E1 and M1 widths obtained from Eqs.~(\ref{eq:E1_width}) and (\ref{eq:M1_width}). The tables include $^3P_1$, $^3D_2$, and radially excited channels for which comparison values are scarce or unavailable.

\subsubsection{E1 transitions}

In Table~\ref{tab:sw_e1}, the largest $S\to P$ widths belong to the adjacent-shell transitions $nS\to(n-1)P$: $4\,^1S_0\to3\,^1P_1$ (49.598~keV), $3\,^1S_0\to2\,^1P_1$ (38.472~keV), $4\,^3S_1\to3\,^3P_1$ (24.111~keV), and $2\,^1S_0\to1\,^1P_1$ (23.308~keV). These channels have photon energies of only 116--142~MeV, yet exceed the transitions from the same initial states to the $1P$ multiplet, whose photon energies are four to seven times larger. Conversely, $3\,^3S_1\to1\,^3P_0$ and $4\,^3S_1\to1\,^3P_0$ have the largest photon energies but among the smallest widths (1.035 and 0.666~keV). This pattern reflects the radial overlap. The $nS$ and $(n-1)P$ wave functions have similar node structure and extent, whereas in cross-shell transitions the sign changes of the integrand suppress $|\epsilon_{fi}|^2$ more strongly than $E_\gamma^3$ can compensate.

For $^3S_1\to{}^3P_J$, Eq.~(\ref{eq:Cfi}) gives the weights $1:3:5$ for $J=0,1,2$. For $2\,^3S_1\to1\,^3P_J$ the widths are 4.897, 10.868, and 8.753~keV. The ratio $\Gamma(1\,^3P_1)/\Gamma(1\,^3P_0)=2.22$ is close to the value 2.19 expected from the angular factor and $E_\gamma^3$ alone, so the radial overlaps are nearly independent of $J$. At large photon energies, as in $4\,^3S_1\to1\,^3P_J$, the ordering $\Gamma(J=2)>\Gamma(J=1)>\Gamma(J=0)$ is recovered, but the ratio $1:2.5:2.9$ is reduced relative to $1:3:5$ by the $E_\gamma$ dependence of the overlap integral of Eq.~(\ref{eq:overlap_E1}).

For the $\Delta n=1$ transitions our results agree reasonably with other approaches. The $2\,^3S_1\to1\,^3P_{0,1}$ widths are close to those of Ref.~\cite{Soni:2017wvy} (4.782 and 11.156~keV) and Ref.~\cite{Ebert:2002pp} (5.53 and 7.65~keV), and the $3\,^3S_1\to2\,^3P_{0,1}$ widths are consistent with Refs.~\cite{Soni:2017wvy,Bokade:2025lmn}. The $2\,^1S_0\to1\,^1P_1$ width lies above the 3--19~keV range of the comparison references, owing to the comparatively low $1\,^1P_1$ mass. For transitions from $n=3,4$ states to the $1P$ multiplet, the model predictions spread over more than an order of magnitude, so these channels may help discriminate among models.

\begin{table*}[htbp]
\centering
\caption{E1 radiative transition widths $\Gamma_{E1}$ (keV) and photon energies $E_\gamma$ (MeV) for $S$-wave transitions of the $B_c$ meson.}
\label{tab:sw_e1}
\setlength{\tabcolsep}{4pt}
\begin{tabular}{llccccccccccc}
\hline\hline
\rule{0pt}{14pt}Initial State  & Final State & $E_\gamma$ (MeV) & This Work &\cite{Bokade:2025lmn} &\cite{Li:2019tbn} & \cite{Li:2023wgq} & \cite{Asghar:2019qjl} & \cite{Akbar:2018hiw} & \cite{Soni:2017wvy} & \cite{Ebert:2002pp} & \cite{Godfrey:2004ya} & \cite{Devlani:2014nda} \\ \\
\hline

\rule{0pt}{12pt}$2^1S_0$ & $1^1P_1$ & 142.2 & 23.308 & 12.215 & 6.38  & 4.4  & 7.11  & 11.59  & 18.663 & 4.40 & 6.1  & 3.03 \\[4pt]
$2^3S_1$ & $1^3P_0$          & 177.0 & 4.897 &  6.631 & 3.48  & 3.0  & 2.42  &  1.395 &  4.782 & 5.53 & 2.9  & 0.94 \\
         & $1^3P_1$          & 159.5 & 10.868 &  ---   & ---   & ---  & ---   &  ---   & 11.156 & 7.65 & 4.7  & 1.45 \\
         & $1^3P_2$          & 124.4 & 8.753 & 10.052 & 6.98  & 5.3  & 4.31  &  2.092 & 16.823 & 7.59 & 5.7  & 2.28 \\[4pt]
$3^1S_0$ & $1^1P_1$          & 532.5 & 5.978 &  1.506 & 1.74  & 3.0  & 0.6   &  1.399 & 38.755 & ---  & ---  & ---  \\
         & $2^1P_1$          & 126.4 & 38.472 & 13.376 & 11.13 & 9.0  & 13.56 &  ---   & 27.988 & ---  & ---  & ---  \\[4pt]
$3^3S_1$ & $1^3P_0$          & 565.0 & 1.035 &  0.601 & 1.73  & 0.04 & 0.09  & 43.877 &  6.910 & ---  & ---  & ---  \\
         & $1^3P_1$          & 548.5 & 2.493 &  ---   & ---   & ---  & ---   &  ---   & 17.563 & ---  & ---  & ---  \\
         & $1^3P_2$          & 515.3 & 2.585 &  0.973 & 1.58  & 2.2  & 0.35  & 161.84 & 27.487 & ---  & ---  & ---  \\
         & $2^3P_0$          & 159.9 & 8.384 &  9.387 & 5.0   & 3.8  & 4.13  &  0.672 &  7.406 & ---  & ---  & ---  \\
         & $2^3P_1$          & 143.1 & 18.345 &  ---   & ---   & ---  & ---   &  ---   & 17.049 & ---  & ---  & ---  \\
         & $2^3P_2$          & 109.3 & 14.055 & 10.854 & 11.89 & 7.6  & 7.09  &  1.713 & 25.112 & ---  & ---  & ---  \\[4pt]
$4^1S_0$ & $1^1P_1$          & 843.4 & 4.257 &  0.192 & 1.93  & ---  & ---   &  ---   &  ---   & ---  & ---  & ---  \\
         & $2^1P_1$          & 455.5 & 17.397 &  8.384 & 6.31   & ---  & ---   &  ---   &  ---   & ---  & ---  & ---  \\
         & $3^1P_1$          & 116.5 & 49.598 &  26.799   & 13.0   & ---  & ---   &  ---   &  ---   & ---  & ---  & ---  \\[4pt]
$4^3S_1$ & $1^3P_0$          & 874.2 & 0.666 &  0.139   & 1.3  & ---  & ---   &  155.86   &  ---   & ---  & ---  & ---  \\
         & $1^3P_1$          & 858.4 & 1.682 &  ---   & ---   & ---  & ---   &  ---   &  ---   & ---  & ---  & ---  \\
         & $1^3P_2$          & 826.8 & 1.951 &  0.112   & 1.88   & ---  & ---   &  648.85   &  ---   & ---  & ---  & ---  \\
         & $2^3P_0$          & 487.3 & 3.046 & 3.581  & 3.28 & ---  & ---   &  24.139   &  ---   & ---  & ---  & ---  \\
         & $2^3P_1$          & 471.3 & 7.310 &  ---   & ---   & ---  & ---   &  ---   &  ---   & ---  & ---  & ---  \\
         & $2^3P_2$          & 439.0 & 7.512 &  5.764   & 5.78  & ---  & ---   & 101.987  &  ---   & ---  & ---  & ---  \\
         & $3^3P_0$          & 149.7 & 11.149 &  22.368 & 6.12  & ---  & ---   &  0.469   &  ---   & ---  & ---  & ---  \\
         & $3^3P_1$          & 133.1 & 24.111 &  ---   & ---   & ---  & ---   &  ---   &  ---   & ---  & ---  & ---  \\
         & $3^3P_2$          & 99.7 & 17.646 &  ---   & ---   & ---  & ---   &  ---   &  ---   & ---  & ---  & ---  \\[4pt]

\hline

\end{tabular}
\end{table*}

In Table~\ref{tab:pw_e1}, the four $1P\to1S$ channels are the largest, with widths of 81--112~keV at photon energies of 407--457~MeV: $1\,^3P_0\to1\,^3S_1$ (81.476~keV), $1\,^3P_1\to1\,^3S_1$ (90.946~keV), $1\,^1P_1\to1\,^1S_0$ (101.166~keV), and $1\,^3P_2\to1\,^3S_1$ (111.554~keV). Since $C_{fi}=1/3$ for all $^3P_J\to{}^3S_1$ transitions, these widths are ordered by the photon energy, and the ratio $\Gamma(1\,^3P_2)/\Gamma(1\,^3P_0)=1.37$ is close to the corresponding $E_\gamma^3$ ratio of 1.42. The $1\,^1P_1$ and $1\,^3P_0$ widths lie within the ranges of Refs.~\cite{Li:2019tbn,Bokade:2025lmn,Li:2022bre,Asghar:2019qjl}, and the $1\,^3P_2$ width is within 4\% of the value 107.402~keV of Ref.~\cite{Bokade:2025lmn}. These channels are directly relevant to the structures observed by LHCb~\cite{LHCb:2025uce}.

For the radially excited $P$-wave states, same-shell transitions are comparable to or larger than cross-shell ones. For example, $2\,^3P_2\to2\,^3S_1$ (75.034~keV, $E_\gamma=293.0$~MeV) exceeds $2\,^3P_2\to1\,^3S_1$ (56.935~keV, $E_\gamma=842.5$~MeV), and $3\,^1P_1\to3\,^1S_0$ (57.842~keV) exceeds both $3\,^1P_1\to2\,^1S_0$ (46.143~keV) and $3\,^1P_1\to1\,^1S_0$ (38.268~keV). The $nP\to n'D$ channels follow the statistical weights closely. For $2\,^3P_2\to1\,^3D_J$ the widths 0.239, 3.564, and 19.799~keV are in the ratio $1:14.9:82.8$, compared with $1:15:84$ from Eq.~(\ref{eq:Cfi}) at nearly equal photon energies, and $2\,^3P_1\to1\,^3D_{1,2}$ follows the expected ratio $1:3$. The largest $2P\to1D$ widths, $2\,^3P_2\to1\,^3D_3$ (19.799~keV) and $2\,^1P_1\to1\,^1D_2$ (17.277~keV), may provide a radiative feeding mechanism for the $1D$ multiplet.

\begin{table*}[htbp]
\centering
\caption{E1 radiative widths $\Gamma_{E1}$ (keV) and photon energies
$E_\gamma$ (MeV) for $P$-wave transitions of the $B_c$ meson.}
\label{tab:pw_e1}
\setlength{\tabcolsep}{10pt}
\begin{tabular}{llcccccc}
\hline\hline
\rule{0pt}{14pt}Initial & Final & $E_\gamma$ & This Work  &\cite{Bokade:2025lmn} & \cite{Li:2022bre}& \cite{Li:2019tbn} & \cite{Asghar:2019qjl} \\[4pt]
\hline
\rule{0pt}{12pt}$1^1P_1$ & $1^1S_0$ & 441.1 & 101.166 & 128.839 & 76.5  & 74.0  & 46.13 \\
$1^3P_0$ & $1^3S_1$          & 407.0 & 81.476 &  68.003 & 59.58 & 96.0  & 52.23 \\
$1^3P_1$ & $1^3S_1$          & 423.9 & 90.946 &  ---    & ---   & ---   & ---   \\
$1^3P_2$ & $1^3S_1$          & 457.4 & 111.554 & 107.402 & 90.04 & 87.0  & 71.91 \\[4pt]
$2^1P_1$ & $1^1S_0$          & 828.2 & 53.318 &  34.326 & 24.42 & 44.0  & 14.64 \\
         & $2^1S_0$          & 276.8 & 64.676 &  64.429 & 43.0  & 36.0  & 24.67 \\
$2^3P_0$ & $1^3S_1$          & 796.9 & 46.354 &  25.434 &  3.19 & 41.0  & 17.52 \\
         & $2^3S_1$          & 243.4 & 45.976 &  32.170 & 33.49 & 53.0  & 27.05 \\
$2^3P_1$ & $1^3S_1$          & 812.1 & 49.710 &  ---    & ---   & ---   & ---   \\
         & $2^3S_1$          & 260.0 & 54.861 &  ---    & ---   & ---   & ---   \\
$2^3P_2$ & $1^3S_1$          & 842.5 & 56.935 &  35.450 & 30.35 & 52.0  & 20.36 \\
         & $2^3S_1$          & 293.0 & 75.034 &  68.728 & 49.37 & 50.0  & 44.67 \\[4pt]
$3^1P_1$ & $1^1S_0$          & 1133.6 & 38.268 &  31.185 & 32.0  & ---  & ---   \\
         & $2^1S_0$          & 608.1 & 46.143 &  54.076 & 28.0  & ---  & ---   \\
         & $3^1S_0$          & 221.7 & 57.842 & 104.341 & 30.0  & ---  & ---   \\
$3^3P_0$ & $1^3S_1$          & 1103.7 & 34.172 &  26.835 & 30.0  & ---  & ---   \\
         & $2^3S_1$          & 576.3 & 37.360 &  37.692 & 36.0  & ---  & ---   \\
         & $3^3S_1$          & 188.6 & 37.927 &  47.035 & 45.0  & ---  & ---   \\
$3^3P_1$ & $1^3S_1$          & 1118.1 & 36.150 &  ---    & ---   & ---  & ---   \\
         & $2^3S_1$          & 592.0 & 41.551 &  ---    & ---   & ---  & ---   \\
         & $3^3S_1$          & 205.1 & 47.343 &  ---    & ---   & ---  & ---   \\
$3^3P_2$ & $1^3S_1$          & 1146.9 & 40.373 &  34.947 & 42.0  & ---  & ---   \\
         & $2^3S_1$          & 623.1 & 50.913 &  62.954 & 39.0  & ---  & ---   \\
         & $3^3S_1$          & 238.0 & 69.168 & 126.179 & 43.0  & ---  & ---   \\[4pt]
\rule{0pt}{12pt}$2^1P_1$ & $1^1D_2$ & 143.3 & 17.277 &  6.297 & 0.96 & ---  & 1.05  \\[4pt]
$2^3P_0$ & $1^3D_1$          & 109.8 & 7.989 &  2.560 & 3.24 & 5.6  & 2.4   \\[4pt]
$2^3P_1$ & $1^3D_1$          & 126.7 & 3.031 &  ---   & ---  & ---  & ---   \\
         & $1^3D_2$          & 126.4 & 9.031 &  ---   & ---  & ---  & ---   \\[4pt]
$2^3P_2$ & $1^3D_1$          & 160.4 & 0.239 &  0.171 & 0.112& ---  & 0.08  \\
         & $1^3D_2$          & 160.1 & 3.564 &  ---   & ---  & ---  & ---   \\
         & $1^3D_3$          & 159.7 & 19.799 & 13.463 & 9.78 & ---  & 5.4   \\[4pt]
$3^1P_1$ & $1^1D_2$          & 480.9 & 7.135 &  0.894 & 0.19 & ---  & ---   \\
         & $2^1D_2$          & 130.2 & 28.393 & 10.143 & 0.19 & ---  & ---   \\[4pt]
$3^3P_0$ & $1^3D_1$          & 449.1 & 4.685 &  1.138 & 1.84 & ---  & ---   \\
         & $2^3D_1$          & 99.7 & 13.209 &  7.994 & 10.93& ---  & ---   \\[4pt]
$3^3P_1$ & $1^3D_1$          & 465.0 & 1.453 &  ---   & ---  & ---  & ---   \\
         & $1^3D_2$          & 464.7 & 4.344 &  ---   & ---  & ---  & ---   \\
         & $2^3D_1$          & 116.4 & 5.162 &  ---   & ---  & ---  & ---   \\
         & $2^3D_2$          & 114.4 & 14.748 &  ---   & ---  & ---  & ---   \\[4pt]
$3^3P_2$ & $1^3D_1$          & 496.8 & 0.0871 &  0.044 & 0.94 & ---  & ---   \\
         & $1^3D_2$          & 496.5 & 1.302 &  ---   & ---  & ---  & ---   \\
         & $1^3D_3$          & 496.0 & 7.255 &  3.517 & 9.07 & ---  & ---   \\
         & $2^3D_1$          & 149.7 & 0.420 &  0.634 & 0.23 & ---  & ---   \\
         & $2^3D_2$          & 147.8 & 6.079 &  ---   & ---  & ---  & ---   \\
         & $2^3D_3$          & 144.8 & 32.190 & 42.453 & 22.0 & ---  & ---   \\[4pt]
\hline\hline
\end{tabular}
\end{table*}

In Table~\ref{tab:dw_e1}, the four $1D\to1P$ channels have widths of 71--93~keV at photon energies of 258--309~MeV: $1\,^1D_2\to1\,^1P_1$ (92.846~keV), $1\,^3D_2\to1\,^3P_1$ (81.987~keV), $1\,^3D_3\to1\,^3P_2$ (77.608~keV), and $1\,^3D_1\to1\,^3P_0$ (70.774~keV). The $1\,^3D_3\to1\,^3P_2$ width is close to the values 78.0 and 86.604~keV of Refs.~\cite{Godfrey:2004ya,Bokade:2025lmn}, and $1\,^3D_1\to1\,^3P_0$ is close to the 65.0~keV of Ref.~\cite{Li:2019tbn}. For $1\,^3D_1\to1\,^3P_J$ the weights $20:15:1$ give 70.774, 45.424, and 2.139~keV. The suppressed $1\,^3D_1\to1\,^3P_2$ width agrees with Refs.~\cite{Soni:2017wvy,Bokade:2025lmn,Li:2022bre} (1.786, 2.330, and 1.74~keV), which suggests a mainly kinematic and angular origin of the suppression. Same-shell transitions again dominate: $2\,^3D_3\to2\,^3P_2$ (63.765~keV) exceeds $2\,^3D_3\to1\,^3P_2$ (34.455~keV), and $3\,^1D_2\to3\,^1P_1$ (71.275~keV) exceeds the transitions to $2\,^1P_1$ and $1\,^1P_1$.

Since the $1D$ states lie below the $BD$ threshold, the sizable $1D\to1P$ widths, together with the $1P\to1S$ widths of Table~\ref{tab:pw_e1}, suggest that the cascade $1D\xrightarrow{\gamma}1P\xrightarrow{\gamma}1S$ could be a promising pathway toward the observation of $D$-wave $B_c$ states. Tables~\ref{tab:sw_e1}--\ref{tab:dw_e1} show the same behavior in all three orbital sectors: the angular dependence follows the E1 statistical weights, while the relative size of the widths is set mainly by the radial overlap.

\begin{table*}[htbp]
\centering
\caption{E1 radiative widths $\Gamma_{E1}$ (keV) and photon energies
$E_\gamma$ (MeV) for $D$-wave transitions of the $B_c$ meson.}
\label{tab:dw_e1}
\setlength{\tabcolsep}{6pt}
\begin{tabular}{llccccccccc}
\hline\hline
\rule{0pt}{14pt}Initial & Final & $E_\gamma$ & This Work & \cite{Bokade:2025lmn} & \cite{Li:2022bre} & \cite{Li:2019tbn} & \cite{Asghar:2019qjl} & \cite{Godfrey:2004ya} & \cite{Soni:2017wvy} & \cite{Ebert:2002pp} \\[4pt]
\hline
\rule{0pt}{12pt}$1^1D_2$ & $1^1P_1$ & 275.7 & 92.846 & 84.679 & 16.3  & 41.0  & 52.26 & 63.0 & 66.020 & 143   \\[4pt]
$1^3D_1$ & $1^3P_0$          & 309.2 & 70.774 & 86.610 & 52.36 & 65.0  & 40.55 & 55.0 & 44.783 & 133   \\
         & $1^3P_1$          & 292.1 & 45.424 &  ---   & ---   & ---   & ---   & ---  & 28.731 & 65.3  \\
         & $1^3P_2$          & 257.6 & 2.139 &  2.330 &  1.74 &  0.7  &  1.2  &  1.8 &  1.786 &  3.82 \\[4pt]
$1^3D_2$ & $1^3P_1$          & 292.4 & 81.987 &  ---   & ---   & ---   & ---   & ---  & 51.272 & 139   \\
         & $1^3P_2$          & 257.9 & 19.310 &  ---   & ---   & ---   & ---   & ---  & 16.073 & 23.6  \\[4pt]
$1^3D_3$ & $1^3P_2$          & 258.3 & 77.608 & 86.604 & 65.98 & 67.0  & 47.81 & 78.0 & 60.336 & 149   \\[4pt]
$2^1D_2$ & $1^1P_1$          & 618.9 & 38.399 & 20.638 &  1.7  & 19.0  &  6.17 & ---  &  ---   & ---   \\
         & $2^1P_1$          & 217.9 & 77.203 & 66.383 &  8.4  & 29.0  & 45.81 & ---  &  ---   & ---   \\[4pt]
$2^3D_1$ & $1^3P_0$          & 648.2 & 26.259 & 20.658 & 13.5  & 41.8  &  3.99 & ---  &  ---   & ---   \\
         & $1^3P_1$          & 631.8 & 17.581 &  ---   & ---   & ---   & ---   & ---  &  ---   & ---   \\
         & $1^3P_2$          & 599.1 & 0.925 &  0.631 &  0.1  &  8.13 &  0.16 & ---  &  ---   & ---   \\
         & $2^3P_0$          & 248.1 & 60.453 & 63.198 & 37.2  & 46.0  & 35.09 & ---  &  ---   & ---   \\
         & $2^3P_1$          & 231.5 & 37.814 &  ---   & ---   & ---   & ---   & ---  &  ---   & ---   \\
         & $2^3P_2$          & 198.1 & 1.658 &  1.378 &  1.29 &  0.58 &  0.95 & ---  &  ---   & ---   \\[4pt]
$2^3D_2$ & $1^3P_1$          & 633.7 & 32.056 &  ---   & ---   & ---   & ---   & ---  &  ---   & ---   \\
         & $1^3P_2$          & 600.9 & 8.441 &  ---   & ---   & ---   & ---   & ---  &  ---   & ---   \\
         & $2^3P_1$          & 233.4 & 69.577 &  ---   & ---   & ---   & ---   & ---  &  ---   & ---   \\
         & $2^3P_2$          & 200.0 & 15.325 &  ---   & ---   & ---   & ---   & ---  &  ---   & ---   \\[4pt]
$2^3D_3$ & $1^3P_2$          & 603.7 & 34.455 & 24.422 &  7.46 & 32.0  &  6.0  & ---  &  ---   & ---   \\
         & $2^3P_2$          & 202.9 & 63.765 & 57.045 & 48.7  & 54.0  & 39.62 & ---  &  ---   & ---   \\[4pt]
$3^1D_2$ & $1^1P_1$          & 907.5 & 24.184 &  ---   & ---   & ---   & ---   & ---  &  ---   & ---   \\
         & $2^1P_1$          & 523.2 & 49.846 &  ---   & ---   & ---   & ---   & ---  &  ---   & ---   \\
         & $3^1P_1$          & 187.4 & 71.275 &  ---   & ---   & ---   & ---   & ---  &  ---   & ---   \\[4pt]
$3^3D_1$ & $1^3P_0$          & 934.0 & 15.789 &  ---   & ---   & ---   & ---   & ---  &  ---   & ---   \\
         & $1^3P_1$          & 918.3 & 10.780 &  ---   & ---   & ---   & ---   & ---  &  ---   & ---   \\
         & $1^3P_2$          & 886.9 & 0.593 &  ---   & ---   & ---   & ---   & ---  &  ---   & ---   \\
         & $2^3P_0$          & 550.5 & 34.793 &  ---   & ---   & ---   & ---   & ---  &  ---   & ---   \\
         & $2^3P_1$          & 534.6 & 22.890 &  ---   & ---   & ---   & ---   & ---  &  ---   & ---   \\
         & $2^3P_2$          & 502.6 & 1.157 &  ---   & ---   & ---   & ---   & ---  &  ---   & ---   \\
         & $3^3P_0$          & 215.9 & 56.811 &  ---   & ---   & ---   & ---   & ---  &  ---   & ---   \\
         & $3^3P_1$          & 199.4 & 34.862 &  ---   & ---   & ---   & ---   & ---  &  ---   & ---   \\
         & $3^3P_2$          & 166.3 & 1.441 &  ---   & ---   & ---   & ---   & ---  &  ---   & ---   \\[4pt]
$3^3D_2$ & $1^3P_1$          & 921.0 & 19.725 &  ---   & ---   & ---   & ---   & ---  &  ---   & ---   \\
         & $1^3P_2$          & 889.6 & 5.426 &  ---   & ---   & ---   & ---   & ---  &  ---   & ---   \\
         & $2^3P_1$          & 537.4 & 42.196 &  ---   & ---   & ---   & ---   & ---  &  ---   & ---   \\
         & $2^3P_2$          & 505.4 & 10.684 &  ---   & ---   & ---   & ---   & ---  &  ---   & ---   \\
         & $3^3P_1$          & 202.4 & 65.186 &  ---   & ---   & ---   & ---   & ---  &  ---   & ---   \\
         & $3^3P_2$          & 169.3 & 13.610 &  ---   & ---   & ---   & ---   & ---  &  ---   & ---   \\[4pt]
$3^3D_3$ & $1^3P_2$          & 893.7 & 22.261 &  ---   & ---   & ---   & ---   & ---  &  ---   & ---   \\
         & $2^3P_2$          & 509.7 & 44.396 &  ---   & ---   & ---   & ---   & ---  &  ---   & ---   \\
         & $3^3P_2$          & 173.8 & 58.405 &  ---   & ---   & ---   & ---   & ---  &  ---   & ---   \\[4pt]
\hline\hline
\end{tabular}
\end{table*}

\subsubsection{M1 transitions}

In Table~\ref{tab:m1_sw}, the allowed transitions ($n=n'$) have $m_{fi}\simeq1$ and photon energies equal to the hyperfine splitting $\Delta M(nS)$, so that $\Gamma_{M1}\propto[\Delta M(nS)]^3$. With the present splittings of 1--2~MeV the allowed widths are below $10^{-2}$~eV. The ATLAS splitting of 64.5~MeV~\cite{ATLAS:2026Bcstar} would instead give $\Gamma_{M1}(1\,^3S_1\to1\,^1S_0)\approx89$~eV, within the 33--150~eV range of most comparison references. These widths should therefore be read together with Sec.~\ref{subsec:massspectrum}.

The hindered transitions ($n\neq n'$) are less sensitive to the hyperfine scale, since their photon energies of several hundred MeV are fixed mainly by the radial spacings. Their widths lie between about 41 and 413~eV and increase with $E_\gamma$ for a given initial state, reaching 412.736~eV for $5\,^1S_0\to1\,^3S_1$. Because the singlet and triplet wave functions are similar in the present framework, the hindered amplitudes arise mainly from the factor $j_0(E_\gamma r/2)$ in Eq.~(\ref{eq:overlap_M1}), which breaks the orthogonality of states with different radial quantum numbers. Transitions connecting the same pair of radial levels then differ mainly through the factor $2J_f+1$. For example, $\Gamma(2\,^1S_0\to1\,^3S_1)/\Gamma(2\,^3S_1\to1\,^1S_0)=2.78$ and $\Gamma(4\,^1S_0\to1\,^3S_1)/\Gamma(4\,^3S_1\to1\,^1S_0)=2.90$, close to the expected value of 3. For $2\,^1S_0\to1\,^3S_1$ we obtain 194.315~eV, close to the 145.843~eV of Ref.~\cite{Bokade:2025lmn} and in the lower part of the 99--1092~eV range of other approaches. The $3\,^1S_0\to2\,^3S_1$ width of 121.740~eV is comparable to the values 62.319 and 152~eV of Refs.~\cite{Bokade:2025lmn,Li:2019tbn}.

\begin{table*}[htbp]
\centering
\caption{M1 radiative widths $\Gamma_{M1}$ (eV) and photon energies
$E_\gamma$ (MeV) for $S$-wave states of the $B_c$ meson.}
\label{tab:m1_sw}
\setlength{\tabcolsep}{3pt}
\begin{tabular}{llccccccccccc}
\hline\hline
\rule{0pt}{14pt}Initial & Final & $E_\gamma$ & This Work &\cite{Bokade:2025lmn} & \cite{Li:2019tbn} &\cite{Li:2023wgq} & \cite{Asghar:2019qjl} & \cite{Godfrey:2004ya} & \cite{Soni:2017wvy} & \cite{Ebert:2002pp}  & \cite{Devlani:2014nda} \\[4pt]
\hline
\rule{0pt}{12pt}$1^3S_1$ & $1^1S_0$ & 1.8 & 0.00194 & 98.422 & 57   & 84   & 150  &  80   &  53.109 &  33  &   2.2  \\[4pt]
$2^1S_0$ & $1^3S_1$          & 572 & 194.315 & 145.843 &  99  & 321  & 340  & 300   & 568.346 & 488  &  1092   \\[4pt]
$2^3S_1$ & $2^1S_0$          & 1.1 & 0.000420 &  20.733 &  2.4 &  8.3 & 0.14 &  10   &  21.119 &  17  &   0.014 \\
         & $1^1S_0$          & 575 & 69.967 & 140.618 & 1205 & 559  & 410  & 600   & 481.572 & 428  &  495   \\[4pt]
$3^1S_0$ & $2^3S_1$          & 397 & 121.740 &  62.319 & 152  &  53  &  70  &   60  &  ---    & ---  & ---     \\
         & $1^3S_1$          & 939 & 288.343 &  79.318 & 510  & 444  & 240  & 4200  &  ---    & ---  & ---     \\[4pt]
$3^3S_1$ & $3^1S_0$          & 0.7 & 0.000103 & 26.2 & 8.227 & 0.8 & 2.2 & 0.042 &  &  ---  & ---   \\
         & $2^1S_0$          & 399 & 45.095 &  60.985 & 356  & 139  &  80  &  200  &  ---    & ---  & ---    \\
         & $1^1S_0$          & 941 & 100.505 &  52.857 & 1885 & 503  & 570  &  600  &  ---    & ---  & ---   \\[4pt]
$4^1S_0$ & $3^3S_1$          & 334 & 114.233 &  36.169 & 186  &  16  & ---  &  ---  &  ---    & ---  & ---    \\
         & $2^3S_1$          & 714 & 207.827 &  41.089 & 579  & 114  & ---  &  ---  &  ---    & ---  & ---    \\
         & $1^3S_1$          & 1231 & 356.153 &  51.119 & 1122 & 452  & ---  &  ---  &  ---    & ---  & ---     \\[4pt]
$4^3S_1$ & $4^1S_0$          & 0.5 & 0.0000391 & 20.8 & 4.123 & 0.35 & 1.1 & ---  &  ---  &  --- & ---  \\
         & $3^1S_0$          & 335 & 41.834 &  35.137 & 252  &  62  & ---  &  ---  &  ---    & ---  & ---     \\
         & $2^1S_0$          & 716 & 74.004 &  27.378 & 806  & 151  & ---  &  ---  &  ---    & ---  & ---     \\
         & $1^1S_0$          & 1233 & 122.690 &  28.596 & 2501 & 443  & ---  &  ---  &  ---    & ---  & ---    \\[4pt]
$5^1S_0$ & $4^3S_1$          & 298 & 114.083 &  23.314 & 209  &  0.6 & ---  &  ---  &  ---    & ---  & ---    \\
         & $3^3S_1$          & 620 & 215.308 &  26.687 & 720  &  46  & ---  &  ---  &  ---    & ---  & ---    \\
         & $2^3S_1$          & 986 & 277.573 &  30.248 & 1260 & 138  & ---  &  ---  &  ---    & ---  & ---    \\
         & $1^3S_1$          & 1483 & 412.736 &  35.435 & 1893 & 432  & ---  &  ---  &  ---    & ---  & ---    \\[4pt]
$5^3S_1$ & $5^1S_0$          & 0.4 & 0.0000203 & 17.2 & 2.324 & 0.18 & 0.6 & ---  &  ---  &  ---  & ---   \\
         & $4^1S_0$          & 299 & 41.147 &  22.327 & 210  &  35  & ---  &  ---  &  ---    & ---  & ---    \\
         & $3^1S_0$          & 621 & 76.635 &  17.734 & 675  &  73  & ---  &  ---  &  ---    & ---  & ---    \\
         & $2^1S_0$          & 987 & 97.152 &  16.635 & 1316 & 146  & ---  &  ---  &  ---    & ---  & ---   \\
         & $1^1S_0$          & 1484 & 141.308 &  17.994 & 3107 & 390  & ---  &  ---  &  ---    & ---  & ---   \\
\hline\hline
\end{tabular}
\end{table*}

\subsection{Regge Trajectories}
\label{subsec:regge_results}

\begin{figure}[htbp]
\centering
\begin{minipage}[t]{0.47\linewidth}
  \centering
  \includegraphics[width=\linewidth]{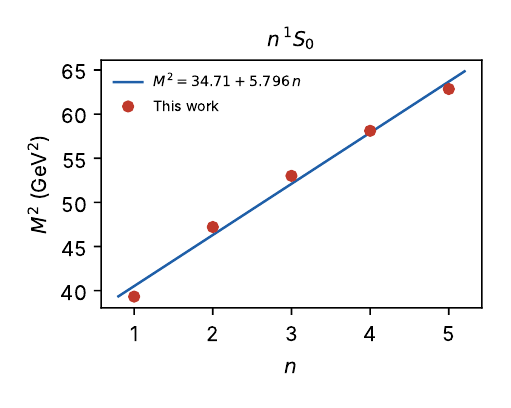}
  \caption{Radial Regge trajectory for the $n\,^1S_0$ $B_c$ states.}
  \label{fig:Regge1}
\end{minipage}
\hfill
\begin{minipage}[t]{0.47\linewidth}
  \centering
  \includegraphics[width=\linewidth]{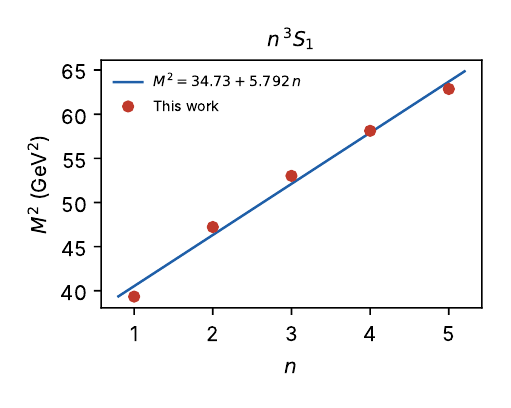}
  \caption{Radial Regge trajectory for the $n\,^3S_1$ $B_c$ states.}
  \label{fig:Regge2}
\end{minipage}
\end{figure}

\begin{figure}[htbp]
\centering
\begin{minipage}[t]{0.47\linewidth}
  \centering
  \includegraphics[width=\linewidth]{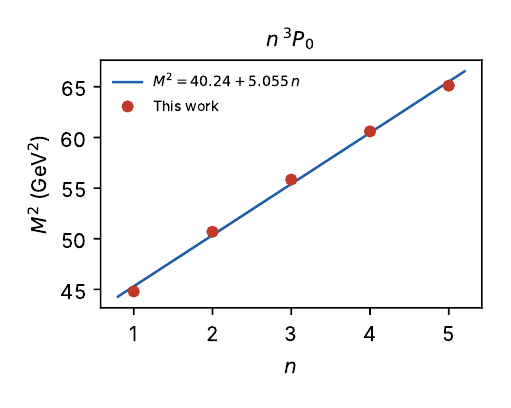}
  \caption{Radial Regge trajectory for the $n\,^3P_0$ $B_c$ states.}
  \label{fig:Regge3}
\end{minipage}
\hfill
\begin{minipage}[t]{0.47\linewidth}
  \centering
  \includegraphics[width=\linewidth]{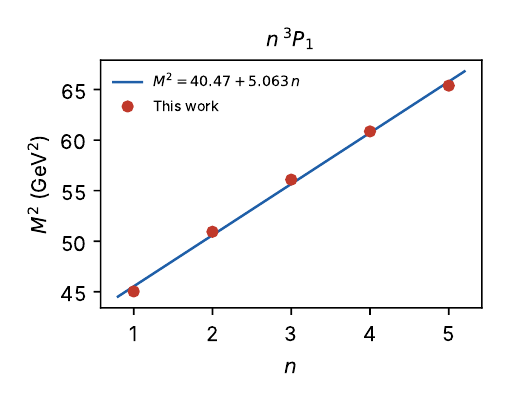}
  \caption{Radial Regge trajectory for the $n\,^3P_1$ $B_c$ states.}
  \label{fig:Regge4}
\end{minipage}
\end{figure}

\begin{figure}[htbp]
\centering
\begin{minipage}[t]{0.47\linewidth}
  \centering
  \includegraphics[width=\linewidth]{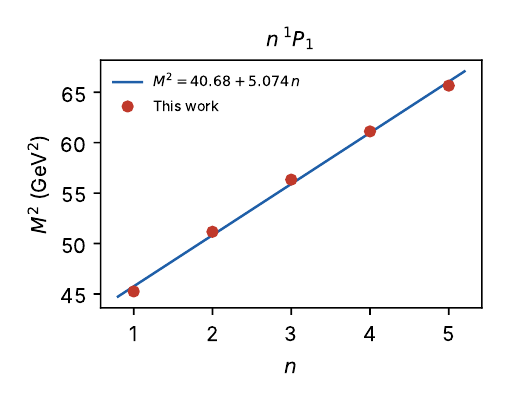}
  \caption{Radial Regge trajectory for the $n\,^1P_1$ $B_c$ states.}
  \label{fig:Regge5}
\end{minipage}
\hfill
\begin{minipage}[t]{0.47\linewidth}
  \centering
  \includegraphics[width=\linewidth]{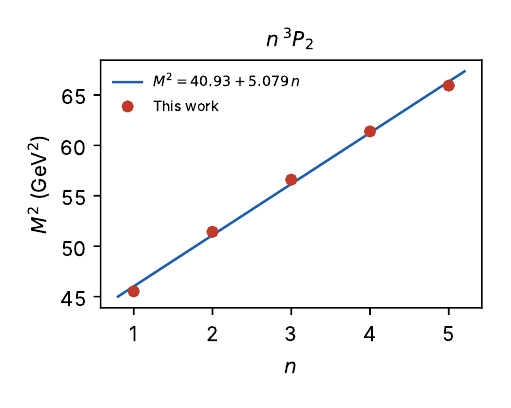}
  \caption{Radial Regge trajectory for the $n\,^3P_2$ $B_c$ states.}
  \label{fig:Regge6}
\end{minipage}
\end{figure}

\begin{figure}[htbp]
\centering
\begin{minipage}[t]{0.47\linewidth}
  \centering
  \includegraphics[width=\linewidth]{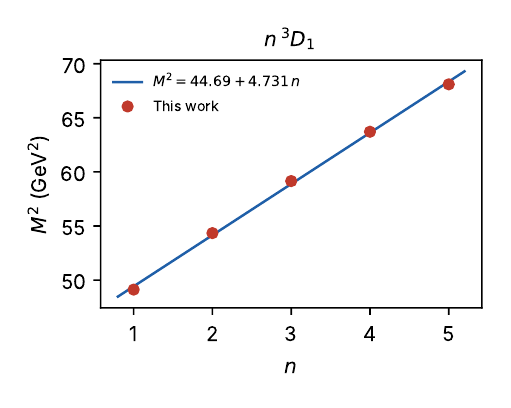}
  \caption{Radial Regge trajectory for the $n\,^3D_1$ $B_c$ states.}
  \label{fig:Regge7}
\end{minipage}
\hfill
\begin{minipage}[t]{0.47\linewidth}
  \centering
  \includegraphics[width=\linewidth]{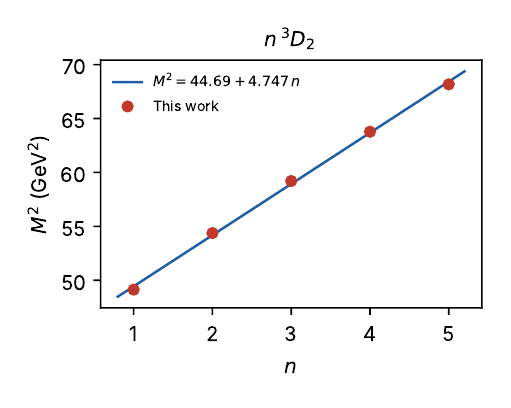}
  \caption{Radial Regge trajectory for the $n\,^3D_2$ $B_c$ states.}
  \label{fig:Regge8}
\end{minipage}
\end{figure}

\begin{figure}[htbp]
\centering
\begin{minipage}[t]{0.47\linewidth}
  \centering
  \includegraphics[width=\linewidth]{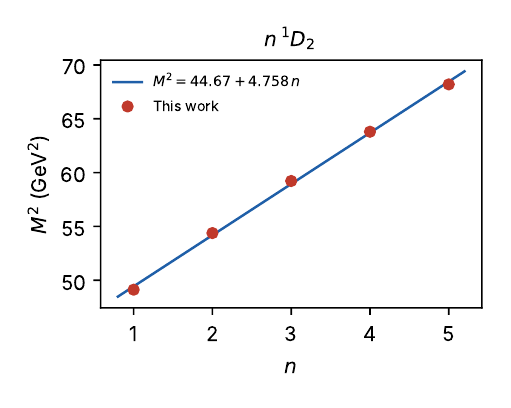}
  \caption{Radial Regge trajectory for the $n\,^1D_2$ $B_c$ states.}
  \label{fig:Regge9}
\end{minipage}
\hfill
\begin{minipage}[t]{0.47\linewidth}
  \centering
  \includegraphics[width=\linewidth]{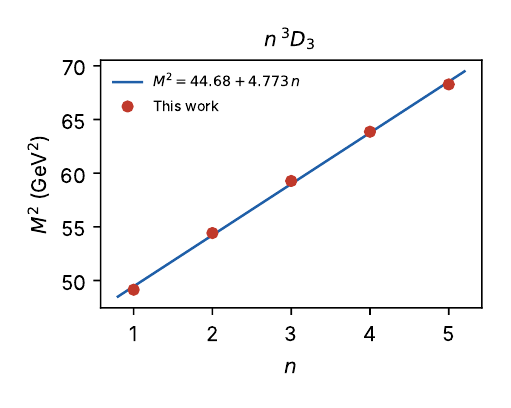}
  \caption{Radial Regge trajectory for the $n\,^3D_3$ $B_c$ states.}
  \label{fig:Regge10}
\end{minipage}
\end{figure}

Figures~\ref{fig:Regge1}--\ref{fig:Regge10} show the radial Regge trajectories obtained from the masses of Tables~\ref{tab:swave}--\ref{tab:dwave_all}, and Table~\ref{tab:regge} lists the fitted parameters of Eq.~(\ref{eq:Regge}). Each trajectory tests the relative spacing of five levels of one spin-parity family at once.

The slopes decrease with increasing orbital angular momentum, from $\beta\approx5.79$~GeV$^2$ for the $S$-wave families to 5.06--5.08~GeV$^2$ for the $P$-wave and 4.73--4.77~GeV$^2$ for the $D$-wave families, a decrease of about 18\%. The intercepts rise from $\beta_0\approx34.7$~GeV$^2$ to 40.2--40.9 and 44.7~GeV$^2$, reflecting the larger masses of the orbitally excited states at fixed $n$. The decrease of $\beta$ follows from the centrifugal barrier, which shifts the wave functions to larger interquark separations. There the linear confining term dominates over the Coulombic one, and the radial excitation energies become more uniform.

Within each orbital sector the trajectories are nearly parallel. The slopes differ by less than 0.01~GeV$^2$ for the two $S$-wave families, by 0.024~GeV$^2$ for the four $P$-wave families, and by 0.042~GeV$^2$ for the four $D$-wave families, i.e., by less than 1\%, which reflects the dominance of the central Cornell interaction in the radial spacing. The small differences follow a regular order: in the $P$-wave sector both $\beta$ and $\beta_0$ increase in the order $^3P_0$, $^3P_1$, $^1P_1$, $^3P_2$, and in the $D$-wave sector $\beta$ increases in the order $^3D_1$, $^3D_2$, $^1D_2$, $^3D_3$. This ordering follows the fine-structure pattern of Sec.~\ref{subsec:massspectrum}, which suggests that the spin--orbit splittings persist along the radial sequence.

The fit quality improves with $L$, from $R^2\approx0.989$ for the $S$-wave to $R^2\approx0.997$ for the $P$-wave and $R^2\approx0.999$ for the $D$-wave families. The deviations from linearity are mildly concave; for the $n\,^1S_0$ family the successive intervals in $M^2$ are 7.88, 5.80, 5.10, and 4.74~GeV$^2$. This curvature is strongest in the $S$-wave sector, where the low-lying states are more sensitive to the Coulombic part of the potential, as also found for triply heavy baryons in the same framework~\cite{Cakir:2026fzd}.

The first $S$-wave interval, $\Delta M^2=7.88$~GeV$^2$, is close to the experimental value of 7.84~GeV$^2$~\cite{ParticleDataGroup:2024cfk}, as expected from the calibration, and the trajectory beyond $n=2$ is a prediction. The same linear fit applied to the $n\,^1S_0$ masses of Ref.~\cite{Ebert:2011jc} gives a slope of 5.77~GeV$^2$, close to the present value, whereas Refs.~\cite{Bokade:2025lmn,Li:2022bre} give about 5.2~GeV$^2$. A measurement of the $3S$ level would directly test the $S$-wave slope.

\clearpage

\begin{table}[h]
\centering
\caption{Radial Regge trajectory parameters for the $B_c$ meson in the $(n,M^2)$ plane, from linear least-squares fits of $M^2=\beta_0+\beta\,n$ [Eq.~(\ref{eq:Regge})] to the masses of Tables~\ref{tab:swave}--\ref{tab:dwave_all}. The intercept $\beta_0$ and slope $\beta$ are in GeV$^2$.}
\label{tab:regge}
\begin{tabular}{lccc}
\hline\hline
Family & $\beta_0$ (GeV$^2$) & $\beta$ (GeV$^2$) & $R^2$ \\
\hline
\multicolumn{4}{c}{$S$-wave} \\
$n\,^1S_0$ & 34.709 & 5.796 & 0.9888 \\
$n\,^3S_1$ & 34.734 & 5.792 & 0.9888 \\
\hline
\multicolumn{4}{c}{$P$-wave} \\
$n\,^3P_0$ & 40.238 & 5.055 & 0.9971 \\
$n\,^3P_1$ & 40.469 & 5.063 & 0.9972 \\
$n\,^1P_1$ & 40.682 & 5.074 & 0.9972 \\
$n\,^3P_2$ & 40.933 & 5.079 & 0.9972 \\
\hline
\multicolumn{4}{c}{$D$-wave} \\
$n\,^3D_1$ & 44.695 & 4.731 & 0.9988 \\
$n\,^3D_2$ & 44.687 & 4.747 & 0.9988 \\
$n\,^1D_2$ & 44.669 & 4.758 & 0.9987 \\
$n\,^3D_3$ & 44.676 & 4.773 & 0.9987 \\
\hline\hline
\end{tabular}
\end{table}

\section{Conclusion}
\label{sec:conclusion}

In this work, we have studied the mass spectrum, leptonic decay constants, E1 and M1 radiative transitions, and radial Regge trajectories of the $B_c$ meson in a nonrelativistic potential model. The Cornell interaction is supplemented by the Breit--Fermi spin--spin and spin--orbit interactions, both included nonperturbatively in the radial Schr\"{o}dinger equation, so that each state has its own wave function and all observables follow from the same set. The parameters, fixed in Ref.~\cite{Cakir:2026fzd} by a fit to the $B_c(1S)$ and $B_c(2S)$ masses, are used without readjustment.

The calibrated $1S$ and $2S$ levels are reproduced. The $1P$ multiplet shows the normal fine-structure ordering, is compatible with lattice QCD, and follows the $1:2$ splitting ratio and the centroid relation expected from the spin--orbit interaction. If the LHCb structures are attributed to the two $1^+$ states, their average mass agrees with our prediction within about 11~MeV, and the larger observed separation may point to $^1P_1$--$^3P_1$ mixing. The $1D$ levels are nearly degenerate below the $BD$ threshold, and all four $D$-wave families are given in the $LS$ basis as input for $2^-$ mixing analyses.

The predicted $f_{B_c}$ of Table~\ref{tab:decay_constant} is consistent with recent QCDSR results but lies above the lattice QCD value, and corresponds to $\mathcal{B}(B_c\to\tau\nu_\tau)\approx3\%$. The ratio $f_{B_c^*}/f_{B_c}$ follows mainly from the one-loop QCD correction.

The E1 widths behave similarly in the $S$-, $P$-, and $D$-wave sectors: their angular dependence follows the statistical weights of the E1 operator, while their relative size is set mainly by the radial overlap. The large $1P\to1S$ and $1D\to1P$ widths suggest that the cascade $1D\xrightarrow{\gamma}1P\xrightarrow{\gamma}1S$ could be a promising route toward the first observation of a $D$-wave $B_c$ state. The hindered M1 widths are determined mainly by the radial spacings and the photon-momentum dependence of the transition operator.

The radial Regge trajectories of all ten spin-parity families are approximately linear and nearly parallel within each orbital sector. Their slopes decrease from the $S$- to the $D$-wave sector as linear confinement becomes dominant at larger interquark separations, and the $S$-wave slope is close to that of the relativistic quark model of Ref.~\cite{Ebert:2011jc}.

Several limitations of the present approach should be acknowledged. First, the tensor interaction is not included, so the fine-structure splittings, in particular the near cancellation in the $D$-wave sector, are approximate. Second, the $S$-wave singlet--triplet splittings of 1--2~MeV lie well below the 64.5~MeV measured by ATLAS. The $S$-wave triplet masses and the allowed M1 widths are therefore the least constrained results, and a joint readjustment of $\sigma$ and the other parameters to this measurement is a natural next step. Third, coupled-channel effects are neglected, so the levels above the $BD$ threshold should be read as bare quark-model states. Finally, $1^+$ and $2^-$ mixing and higher-order QCD corrections to the decay constants are left for future work.

The high-luminosity LHC and future $e^+e^-$ facilities such as FCC-ee~\cite{FCC:2018byv} and CEPC~\cite{CEPCStudyGroup:2018ghi} are expected to provide much larger $B_c$ samples. Measurements of the spin-parity of the $1P$ structures, the $2S$ hyperfine splitting, the $1P\to1S$ and $1D\to1P$ radiative branching fractions, and the decay $B_c\to\tau\nu_\tau$ would test the predictions presented here.

\begin{acknowledgments}
This work was supported by the Scientific Research Projects
Coordination Unit (BAP) of Ondokuz May{\i}s University under
Project No.\ BAP01-2025-5761.
\end{acknowledgments}

\bibliographystyle{apsrev4-2}
\bibliography{Bc_meson_final_v2}

\begin{thebibliography}{83}%
\makeatletter
\providecommand \@ifxundefined [1]{%
 \@ifx{#1\undefined}
}%
\providecommand \@ifnum [1]{%
 \ifnum #1\expandafter \@firstoftwo
 \else \expandafter \@secondoftwo
 \fi
}%
\providecommand \@ifx [1]{%
 \ifx #1\expandafter \@firstoftwo
 \else \expandafter \@secondoftwo
 \fi
}%
\providecommand \natexlab [1]{#1}%
\providecommand \enquote  [1]{``#1''}%
\providecommand \bibnamefont  [1]{#1}%
\providecommand \bibfnamefont [1]{#1}%
\providecommand \citenamefont [1]{#1}%
\providecommand \href@noop [0]{\@secondoftwo}%
\providecommand \href [0]{\begingroup \@sanitize@url \@href}%
\providecommand \@href[1]{\@@startlink{#1}\@@href}%
\providecommand \@@href[1]{\endgroup#1\@@endlink}%
\providecommand \@sanitize@url [0]{\catcode `\\12\catcode `\$12\catcode
  `\&12\catcode `\#12\catcode `\^12\catcode `\_12\catcode `\%12\relax}%
\providecommand \@@startlink[1]{}%
\providecommand \@@endlink[0]{}%
\providecommand \url  [0]{\begingroup\@sanitize@url \@url }%
\providecommand \@url [1]{\endgroup\@href {#1}{\urlprefix }}%
\providecommand \urlprefix  [0]{URL }%
\providecommand \Eprint [0]{\href }%
\providecommand \doibase [0]{https://doi.org/}%
\providecommand \selectlanguage [0]{\@gobble}%
\providecommand \bibinfo  [0]{\@secondoftwo}%
\providecommand \bibfield  [0]{\@secondoftwo}%
\providecommand \translation [1]{[#1]}%
\providecommand \BibitemOpen [0]{}%
\providecommand \bibitemStop [0]{}%
\providecommand \bibitemNoStop [0]{.\EOS\space}%
\providecommand \EOS [0]{\spacefactor3000\relax}%
\providecommand \BibitemShut  [1]{\csname bibitem#1\endcsname}%
\let\auto@bib@innerbib\@empty
\bibitem [{\citenamefont {Alonso}\ \emph {et~al.}(2017)\citenamefont {Alonso},
  \citenamefont {Grinstein},\ and\ \citenamefont
  {Martin~Camalich}}]{Alonso:2016oyd}%
  \BibitemOpen
  \bibfield  {author} {\bibinfo {author} {\bibfnamefont {R.}~\bibnamefont
  {Alonso}}, \bibinfo {author} {\bibfnamefont {B.}~\bibnamefont {Grinstein}},\
  and\ \bibinfo {author} {\bibfnamefont {J.}~\bibnamefont {Martin~Camalich}},\
  }\href {https://doi.org/10.1103/PhysRevLett.118.081802} {\bibfield  {journal}
  {\bibinfo  {journal} {Phys. Rev. Lett.}\ }\textbf {\bibinfo {volume} {118}},\
  \bibinfo {pages} {081802} (\bibinfo {year} {2017})},\ \Eprint
  {https://arxiv.org/abs/1611.06676} {arXiv:1611.06676 [hep-ph]} \BibitemShut
  {NoStop}%
\bibitem [{\citenamefont {Akeroyd}\ and\ \citenamefont
  {Chen}(2017)}]{Akeroyd:2017mhr}%
  \BibitemOpen
  \bibfield  {author} {\bibinfo {author} {\bibfnamefont {A.~G.}\ \bibnamefont
  {Akeroyd}}\ and\ \bibinfo {author} {\bibfnamefont {C.-H.}\ \bibnamefont
  {Chen}},\ }\href {https://doi.org/10.1103/PhysRevD.96.075011} {\bibfield
  {journal} {\bibinfo  {journal} {Phys. Rev. D}\ }\textbf {\bibinfo {volume}
  {96}},\ \bibinfo {pages} {075011} (\bibinfo {year} {2017})},\ \Eprint
  {https://arxiv.org/abs/1708.04072} {arXiv:1708.04072 [hep-ph]} \BibitemShut
  {NoStop}%
\bibitem [{\citenamefont {Costa~i Reina}\ \emph {et~al.}(2026)\citenamefont
  {Costa~i Reina}, \citenamefont {Gubernari},\ and\ \citenamefont
  {W{\"u}thrich}}]{CostaiReina:2026deh}%
  \BibitemOpen
  \bibfield  {author} {\bibinfo {author} {\bibfnamefont {E.}~\bibnamefont
  {Costa~i Reina}}, \bibinfo {author} {\bibfnamefont {N.}~\bibnamefont
  {Gubernari}},\ and\ \bibinfo {author} {\bibfnamefont {Z.}~\bibnamefont
  {W{\"u}thrich}},\ }\href@noop {} {\  (\bibinfo {year} {2026})},\ \Eprint
  {https://arxiv.org/abs/2606.25733} {arXiv:2606.25733 [hep-ph]} \BibitemShut
  {NoStop}%
\bibitem [{\citenamefont {Abe}\ \emph {et~al.}(1998)\citenamefont {Abe} \emph
  {et~al.}}]{Abe:1998wi}%
  \BibitemOpen
  \bibfield  {author} {\bibinfo {author} {\bibfnamefont {F.}~\bibnamefont
  {Abe}} \emph {et~al.} (\bibinfo {collaboration} {CDF}),\ }\href
  {https://doi.org/10.1103/PhysRevLett.81.2432} {\bibfield  {journal} {\bibinfo
   {journal} {Phys. Rev. Lett.}\ }\textbf {\bibinfo {volume} {81}},\ \bibinfo
  {pages} {2432} (\bibinfo {year} {1998})},\ \Eprint
  {https://arxiv.org/abs/hep-ex/9805034} {arXiv:hep-ex/9805034} \BibitemShut
  {NoStop}%
\bibitem [{\citenamefont {Eichten}\ and\ \citenamefont
  {Quigg}(1994)}]{Eichten:1994gt}%
  \BibitemOpen
  \bibfield  {author} {\bibinfo {author} {\bibfnamefont {E.~J.}\ \bibnamefont
  {Eichten}}\ and\ \bibinfo {author} {\bibfnamefont {C.}~\bibnamefont
  {Quigg}},\ }\href {https://doi.org/10.1103/PhysRevD.49.5845} {\bibfield
  {journal} {\bibinfo  {journal} {Phys. Rev. D}\ }\textbf {\bibinfo {volume}
  {49}},\ \bibinfo {pages} {5845} (\bibinfo {year} {1994})},\ \Eprint
  {https://arxiv.org/abs/hep-ph/9402210} {arXiv:hep-ph/9402210} \BibitemShut
  {NoStop}%
\bibitem [{\citenamefont {Gershtein}\ \emph
  {et~al.}(1995{\natexlab{a}})\citenamefont {Gershtein}, \citenamefont
  {Kiselev}, \citenamefont {Likhoded},\ and\ \citenamefont
  {Tkabladze}}]{Gershtein:1994dxw}%
  \BibitemOpen
  \bibfield  {author} {\bibinfo {author} {\bibfnamefont {S.~S.}\ \bibnamefont
  {Gershtein}}, \bibinfo {author} {\bibfnamefont {V.~V.}\ \bibnamefont
  {Kiselev}}, \bibinfo {author} {\bibfnamefont {A.~K.}\ \bibnamefont
  {Likhoded}},\ and\ \bibinfo {author} {\bibfnamefont {A.~V.}\ \bibnamefont
  {Tkabladze}},\ }\href {https://doi.org/10.1103/PhysRevD.51.3613} {\bibfield
  {journal} {\bibinfo  {journal} {Phys. Rev. D}\ }\textbf {\bibinfo {volume}
  {51}},\ \bibinfo {pages} {3613} (\bibinfo {year} {1995}{\natexlab{a}})},\
  \Eprint {https://arxiv.org/abs/hep-ph/9406339} {arXiv:hep-ph/9406339}
  \BibitemShut {NoStop}%
\bibitem [{\citenamefont {Aaij}\ \emph {et~al.}(2012)\citenamefont {Aaij} \emph
  {et~al.}}]{Aaij:2012dd}%
  \BibitemOpen
  \bibfield  {author} {\bibinfo {author} {\bibfnamefont {R.}~\bibnamefont
  {Aaij}} \emph {et~al.} (\bibinfo {collaboration} {LHCb}),\ }\href
  {https://doi.org/10.1103/PhysRevLett.109.232001} {\bibfield  {journal}
  {\bibinfo  {journal} {Phys. Rev. Lett.}\ }\textbf {\bibinfo {volume} {109}},\
  \bibinfo {pages} {232001} (\bibinfo {year} {2012})},\ \Eprint
  {https://arxiv.org/abs/1209.5634} {arXiv:1209.5634 [hep-ex]} \BibitemShut
  {NoStop}%
\bibitem [{\citenamefont {Navas}\ \emph {et~al.}(2024)\citenamefont {Navas}
  \emph {et~al.}}]{ParticleDataGroup:2024cfk}%
  \BibitemOpen
  \bibfield  {author} {\bibinfo {author} {\bibfnamefont {S.}~\bibnamefont
  {Navas}} \emph {et~al.} (\bibinfo {collaboration} {Particle Data Group}),\
  }\href {https://doi.org/10.1103/PhysRevD.110.030001} {\bibfield  {journal}
  {\bibinfo  {journal} {Phys. Rev. D}\ }\textbf {\bibinfo {volume} {110}},\
  \bibinfo {pages} {030001} (\bibinfo {year} {2024})}\BibitemShut {NoStop}%
\bibitem [{\citenamefont {Aad}\ \emph {et~al.}(2014)\citenamefont {Aad} \emph
  {et~al.}}]{Aad:2014laa}%
  \BibitemOpen
  \bibfield  {author} {\bibinfo {author} {\bibfnamefont {G.}~\bibnamefont
  {Aad}} \emph {et~al.} (\bibinfo {collaboration} {ATLAS}),\ }\href
  {https://doi.org/10.1103/PhysRevLett.113.212004} {\bibfield  {journal}
  {\bibinfo  {journal} {Phys. Rev. Lett.}\ }\textbf {\bibinfo {volume} {113}},\
  \bibinfo {pages} {212004} (\bibinfo {year} {2014})},\ \Eprint
  {https://arxiv.org/abs/1407.1032} {arXiv:1407.1032 [hep-ex]} \BibitemShut
  {NoStop}%
\bibitem [{\citenamefont {Sirunyan}\ \emph {et~al.}(2019)\citenamefont
  {Sirunyan} \emph {et~al.}}]{CMS:2019uhm}%
  \BibitemOpen
  \bibfield  {author} {\bibinfo {author} {\bibfnamefont {A.~M.}\ \bibnamefont
  {Sirunyan}} \emph {et~al.} (\bibinfo {collaboration} {CMS}),\ }\href
  {https://doi.org/10.1103/PhysRevLett.122.132001} {\bibfield  {journal}
  {\bibinfo  {journal} {Phys. Rev. Lett.}\ }\textbf {\bibinfo {volume} {122}},\
  \bibinfo {pages} {132001} (\bibinfo {year} {2019})},\ \Eprint
  {https://arxiv.org/abs/1902.00571} {arXiv:1902.00571 [hep-ex]} \BibitemShut
  {NoStop}%
\bibitem [{\citenamefont {Aaij}\ \emph {et~al.}(2019)\citenamefont {Aaij} \emph
  {et~al.}}]{LHCb:2019bem}%
  \BibitemOpen
  \bibfield  {author} {\bibinfo {author} {\bibfnamefont {R.}~\bibnamefont
  {Aaij}} \emph {et~al.} (\bibinfo {collaboration} {LHCb}),\ }\href
  {https://doi.org/10.1103/PhysRevLett.122.232001} {\bibfield  {journal}
  {\bibinfo  {journal} {Phys. Rev. Lett.}\ }\textbf {\bibinfo {volume} {122}},\
  \bibinfo {pages} {232001} (\bibinfo {year} {2019})},\ \Eprint
  {https://arxiv.org/abs/1904.00081} {arXiv:1904.00081 [hep-ex]} \BibitemShut
  {NoStop}%
\bibitem [{\citenamefont {Aad}\ \emph {et~al.}(2026)\citenamefont {Aad} \emph
  {et~al.}}]{ATLAS:2026Bcstar}%
  \BibitemOpen
  \bibfield  {author} {\bibinfo {author} {\bibfnamefont {G.}~\bibnamefont
  {Aad}} \emph {et~al.} (\bibinfo {collaboration} {ATLAS}),\ }\href
  {https://doi.org/10.1103/vrxp-zkbs} {\bibfield  {journal} {\bibinfo
  {journal} {Phys. Rev. Lett.}\ }\textbf {\bibinfo {volume} {137}},\ \bibinfo
  {pages} {141901} (\bibinfo {year} {2026})},\ \Eprint
  {https://arxiv.org/abs/2605.16228} {arXiv:2605.16228 [hep-ex]} \BibitemShut
  {NoStop}%
\bibitem [{\citenamefont {Aaij}\ \emph {et~al.}(2025)\citenamefont {Aaij} \emph
  {et~al.}}]{LHCb:2025uce}%
  \BibitemOpen
  \bibfield  {author} {\bibinfo {author} {\bibfnamefont {R.}~\bibnamefont
  {Aaij}} \emph {et~al.} (\bibinfo {collaboration} {LHCb}),\ }\href
  {https://doi.org/10.1103/fc8j-tb8k} {\bibfield  {journal} {\bibinfo
  {journal} {Phys. Rev. Lett.}\ }\textbf {\bibinfo {volume} {135}},\ \bibinfo
  {pages} {231902} (\bibinfo {year} {2025})},\ \Eprint
  {https://arxiv.org/abs/2507.02149} {arXiv:2507.02149 [hep-ex]} \BibitemShut
  {NoStop}%
\bibitem [{\citenamefont {Abada}\ \emph {et~al.}(2019)\citenamefont {Abada}
  \emph {et~al.}}]{FCC:2018byv}%
  \BibitemOpen
  \bibfield  {author} {\bibinfo {author} {\bibfnamefont {A.}~\bibnamefont
  {Abada}} \emph {et~al.} (\bibinfo {collaboration} {FCC}),\ }\href
  {https://doi.org/10.1140/epjc/s10052-019-6904-3} {\bibfield  {journal}
  {\bibinfo  {journal} {Eur. Phys. J. C}\ }\textbf {\bibinfo {volume} {79}},\
  \bibinfo {pages} {474} (\bibinfo {year} {2019})}\BibitemShut {NoStop}%
\bibitem [{\citenamefont {Dong}\ \emph {et~al.}(2018)\citenamefont {Dong} \emph
  {et~al.}}]{CEPCStudyGroup:2018ghi}%
  \BibitemOpen
  \bibfield  {author} {\bibinfo {author} {\bibfnamefont {M.}~\bibnamefont
  {Dong}} \emph {et~al.} (\bibinfo {collaboration} {CEPC Study Group}),\
  }\href@noop {} {\  (\bibinfo {year} {2018})},\ \Eprint
  {https://arxiv.org/abs/1811.10545} {arXiv:1811.10545 [hep-ex]} \BibitemShut
  {NoStop}%
\bibitem [{\citenamefont {Godfrey}\ and\ \citenamefont
  {Isgur}(1985)}]{Godfrey:1985xj}%
  \BibitemOpen
  \bibfield  {author} {\bibinfo {author} {\bibfnamefont {S.}~\bibnamefont
  {Godfrey}}\ and\ \bibinfo {author} {\bibfnamefont {N.}~\bibnamefont
  {Isgur}},\ }\href {https://doi.org/10.1103/PhysRevD.32.189} {\bibfield
  {journal} {\bibinfo  {journal} {Phys. Rev. D}\ }\textbf {\bibinfo {volume}
  {32}},\ \bibinfo {pages} {189} (\bibinfo {year} {1985})}\BibitemShut
  {NoStop}%
\bibitem [{\citenamefont {Zeng}\ \emph {et~al.}(1995)\citenamefont {Zeng},
  \citenamefont {Van~Orden},\ and\ \citenamefont {Roberts}}]{Zeng:1994vj}%
  \BibitemOpen
  \bibfield  {author} {\bibinfo {author} {\bibfnamefont {J.}~\bibnamefont
  {Zeng}}, \bibinfo {author} {\bibfnamefont {J.~W.}\ \bibnamefont
  {Van~Orden}},\ and\ \bibinfo {author} {\bibfnamefont {W.}~\bibnamefont
  {Roberts}},\ }\href {https://doi.org/10.1103/PhysRevD.52.5229} {\bibfield
  {journal} {\bibinfo  {journal} {Phys. Rev. D}\ }\textbf {\bibinfo {volume}
  {52}},\ \bibinfo {pages} {5229} (\bibinfo {year} {1995})},\ \Eprint
  {https://arxiv.org/abs/hep-ph/9412269} {arXiv:hep-ph/9412269} \BibitemShut
  {NoStop}%
\bibitem [{\citenamefont {Gupta}\ and\ \citenamefont
  {Johnson}(1996)}]{Gupta:1995ps}%
  \BibitemOpen
  \bibfield  {author} {\bibinfo {author} {\bibfnamefont {S.~N.}\ \bibnamefont
  {Gupta}}\ and\ \bibinfo {author} {\bibfnamefont {J.~M.}\ \bibnamefont
  {Johnson}},\ }\href {https://doi.org/10.1103/PhysRevD.53.312} {\bibfield
  {journal} {\bibinfo  {journal} {Phys. Rev. D}\ }\textbf {\bibinfo {volume}
  {53}},\ \bibinfo {pages} {312} (\bibinfo {year} {1996})},\ \Eprint
  {https://arxiv.org/abs/hep-ph/9511267} {arXiv:hep-ph/9511267} \BibitemShut
  {NoStop}%
\bibitem [{\citenamefont {Ebert}\ \emph {et~al.}(2003)\citenamefont {Ebert},
  \citenamefont {Faustov},\ and\ \citenamefont {Galkin}}]{Ebert:2002pp}%
  \BibitemOpen
  \bibfield  {author} {\bibinfo {author} {\bibfnamefont {D.}~\bibnamefont
  {Ebert}}, \bibinfo {author} {\bibfnamefont {R.~N.}\ \bibnamefont {Faustov}},\
  and\ \bibinfo {author} {\bibfnamefont {V.~O.}\ \bibnamefont {Galkin}},\
  }\href {https://doi.org/10.1103/PhysRevD.67.014027} {\bibfield  {journal}
  {\bibinfo  {journal} {Phys. Rev. D}\ }\textbf {\bibinfo {volume} {67}},\
  \bibinfo {pages} {014027} (\bibinfo {year} {2003})},\ \Eprint
  {https://arxiv.org/abs/hep-ph/0210381} {arXiv:hep-ph/0210381} \BibitemShut
  {NoStop}%
\bibitem [{\citenamefont {Godfrey}(2004)}]{Godfrey:2004ya}%
  \BibitemOpen
  \bibfield  {author} {\bibinfo {author} {\bibfnamefont {S.}~\bibnamefont
  {Godfrey}},\ }\href {https://doi.org/10.1103/PhysRevD.70.054017} {\bibfield
  {journal} {\bibinfo  {journal} {Phys. Rev. D}\ }\textbf {\bibinfo {volume}
  {70}},\ \bibinfo {pages} {054017} (\bibinfo {year} {2004})},\ \Eprint
  {https://arxiv.org/abs/hep-ph/0406228} {arXiv:hep-ph/0406228} \BibitemShut
  {NoStop}%
\bibitem [{\citenamefont {Fulcher}(1999)}]{Fulcher:1998ka}%
  \BibitemOpen
  \bibfield  {author} {\bibinfo {author} {\bibfnamefont {L.~P.}\ \bibnamefont
  {Fulcher}},\ }\href {https://doi.org/10.1103/PhysRevD.60.074006} {\bibfield
  {journal} {\bibinfo  {journal} {Phys. Rev. D}\ }\textbf {\bibinfo {volume}
  {60}},\ \bibinfo {pages} {074006} (\bibinfo {year} {1999})},\ \Eprint
  {https://arxiv.org/abs/hep-ph/9806444} {arXiv:hep-ph/9806444} \BibitemShut
  {NoStop}%
\bibitem [{\citenamefont {Ebert}\ \emph {et~al.}(2011)\citenamefont {Ebert},
  \citenamefont {Faustov},\ and\ \citenamefont {Galkin}}]{Ebert:2011jc}%
  \BibitemOpen
  \bibfield  {author} {\bibinfo {author} {\bibfnamefont {D.}~\bibnamefont
  {Ebert}}, \bibinfo {author} {\bibfnamefont {R.~N.}\ \bibnamefont {Faustov}},\
  and\ \bibinfo {author} {\bibfnamefont {V.~O.}\ \bibnamefont {Galkin}},\
  }\href {https://doi.org/10.1140/epjc/s10052-011-1825-9} {\bibfield  {journal}
  {\bibinfo  {journal} {Eur. Phys. J. C}\ }\textbf {\bibinfo {volume} {71}},\
  \bibinfo {pages} {1825} (\bibinfo {year} {2011})},\ \Eprint
  {https://arxiv.org/abs/1111.0454} {arXiv:1111.0454 [hep-ph]} \BibitemShut
  {NoStop}%
\bibitem [{\citenamefont {Monteiro}\ \emph {et~al.}(2017)\citenamefont
  {Monteiro}, \citenamefont {Bhat},\ and\ \citenamefont
  {Vijaya~Kumar}}]{Monteiro:2016ijw}%
  \BibitemOpen
  \bibfield  {author} {\bibinfo {author} {\bibfnamefont {A.~P.}\ \bibnamefont
  {Monteiro}}, \bibinfo {author} {\bibfnamefont {M.}~\bibnamefont {Bhat}},\
  and\ \bibinfo {author} {\bibfnamefont {K.~B.}\ \bibnamefont {Vijaya~Kumar}},\
  }\href {https://doi.org/10.1142/S0217751X1750021X} {\bibfield  {journal}
  {\bibinfo  {journal} {Int. J. Mod. Phys. A}\ }\textbf {\bibinfo {volume}
  {32}},\ \bibinfo {pages} {1750021} (\bibinfo {year} {2017})},\ \Eprint
  {https://arxiv.org/abs/1607.07594} {arXiv:1607.07594 [hep-ph]} \BibitemShut
  {NoStop}%
\bibitem [{\citenamefont {Soni}\ \emph {et~al.}(2018)\citenamefont {Soni},
  \citenamefont {Joshi}, \citenamefont {Shah}, \citenamefont {Chauhan},\ and\
  \citenamefont {Pandya}}]{Soni:2017wvy}%
  \BibitemOpen
  \bibfield  {author} {\bibinfo {author} {\bibfnamefont {N.~R.}\ \bibnamefont
  {Soni}}, \bibinfo {author} {\bibfnamefont {B.~R.}\ \bibnamefont {Joshi}},
  \bibinfo {author} {\bibfnamefont {R.~P.}\ \bibnamefont {Shah}}, \bibinfo
  {author} {\bibfnamefont {H.~R.}\ \bibnamefont {Chauhan}},\ and\ \bibinfo
  {author} {\bibfnamefont {J.~N.}\ \bibnamefont {Pandya}},\ }\href
  {https://doi.org/10.1140/epjc/s10052-018-6068-6} {\bibfield  {journal}
  {\bibinfo  {journal} {Eur. Phys. J. C}\ }\textbf {\bibinfo {volume} {78}},\
  \bibinfo {pages} {592} (\bibinfo {year} {2018})},\ \Eprint
  {https://arxiv.org/abs/1707.07144} {arXiv:1707.07144 [hep-ph]} \BibitemShut
  {NoStop}%
\bibitem [{\citenamefont {Eichten}\ and\ \citenamefont
  {Quigg}(2019)}]{Eichten:2019gig}%
  \BibitemOpen
  \bibfield  {author} {\bibinfo {author} {\bibfnamefont {E.~J.}\ \bibnamefont
  {Eichten}}\ and\ \bibinfo {author} {\bibfnamefont {C.}~\bibnamefont
  {Quigg}},\ }\href {https://doi.org/10.1103/PhysRevD.99.054025} {\bibfield
  {journal} {\bibinfo  {journal} {Phys. Rev. D}\ }\textbf {\bibinfo {volume}
  {99}},\ \bibinfo {pages} {054025} (\bibinfo {year} {2019})},\ \Eprint
  {https://arxiv.org/abs/1902.09735} {arXiv:1902.09735 [hep-ph]} \BibitemShut
  {NoStop}%
\bibitem [{\citenamefont {Li}\ \emph {et~al.}(2019)\citenamefont {Li},
  \citenamefont {Liu}, \citenamefont {Lu}, \citenamefont {L{\"u}},
  \citenamefont {Gui},\ and\ \citenamefont {Zhong}}]{Li:2019tbn}%
  \BibitemOpen
  \bibfield  {author} {\bibinfo {author} {\bibfnamefont {Q.}~\bibnamefont
  {Li}}, \bibinfo {author} {\bibfnamefont {M.-S.}\ \bibnamefont {Liu}},
  \bibinfo {author} {\bibfnamefont {L.-S.}\ \bibnamefont {Lu}}, \bibinfo
  {author} {\bibfnamefont {Q.-F.}\ \bibnamefont {L{\"u}}}, \bibinfo {author}
  {\bibfnamefont {L.-C.}\ \bibnamefont {Gui}},\ and\ \bibinfo {author}
  {\bibfnamefont {X.-H.}\ \bibnamefont {Zhong}},\ }\href
  {https://doi.org/10.1103/PhysRevD.99.096020} {\bibfield  {journal} {\bibinfo
  {journal} {Phys. Rev. D}\ }\textbf {\bibinfo {volume} {99}},\ \bibinfo
  {pages} {096020} (\bibinfo {year} {2019})},\ \Eprint
  {https://arxiv.org/abs/1903.11927} {arXiv:1903.11927 [hep-ph]} \BibitemShut
  {NoStop}%
\bibitem [{\citenamefont {Ortega}\ \emph {et~al.}(2020)\citenamefont {Ortega},
  \citenamefont {Segovia}, \citenamefont {Entem},\ and\ \citenamefont
  {Fernandez}}]{Ortega:2020uvc}%
  \BibitemOpen
  \bibfield  {author} {\bibinfo {author} {\bibfnamefont {P.~G.}\ \bibnamefont
  {Ortega}}, \bibinfo {author} {\bibfnamefont {J.}~\bibnamefont {Segovia}},
  \bibinfo {author} {\bibfnamefont {D.~R.}\ \bibnamefont {Entem}},\ and\
  \bibinfo {author} {\bibfnamefont {F.}~\bibnamefont {Fernandez}},\ }\href
  {https://doi.org/10.1140/epjc/s10052-020-7764-6} {\bibfield  {journal}
  {\bibinfo  {journal} {Eur. Phys. J. C}\ }\textbf {\bibinfo {volume} {80}},\
  \bibinfo {pages} {223} (\bibinfo {year} {2020})},\ \Eprint
  {https://arxiv.org/abs/2001.08093} {arXiv:2001.08093 [hep-ph]} \BibitemShut
  {NoStop}%
\bibitem [{\citenamefont {Verma}(2012)}]{Verma:2011yw}%
  \BibitemOpen
  \bibfield  {author} {\bibinfo {author} {\bibfnamefont {R.~C.}\ \bibnamefont
  {Verma}},\ }\href {https://doi.org/10.1088/0954-3899/39/2/025005} {\bibfield
  {journal} {\bibinfo  {journal} {J. Phys. G}\ }\textbf {\bibinfo {volume}
  {39}},\ \bibinfo {pages} {025005} (\bibinfo {year} {2012})},\ \Eprint
  {https://arxiv.org/abs/1103.2973} {arXiv:1103.2973 [hep-ph]} \BibitemShut
  {NoStop}%
\bibitem [{\citenamefont {Tang}\ \emph {et~al.}(2018)\citenamefont {Tang},
  \citenamefont {Li}, \citenamefont {Maris},\ and\ \citenamefont
  {Vary}}]{Tang:2018myz}%
  \BibitemOpen
  \bibfield  {author} {\bibinfo {author} {\bibfnamefont {S.}~\bibnamefont
  {Tang}}, \bibinfo {author} {\bibfnamefont {Y.}~\bibnamefont {Li}}, \bibinfo
  {author} {\bibfnamefont {P.}~\bibnamefont {Maris}},\ and\ \bibinfo {author}
  {\bibfnamefont {J.~P.}\ \bibnamefont {Vary}},\ }\href
  {https://doi.org/10.1103/PhysRevD.98.114038} {\bibfield  {journal} {\bibinfo
  {journal} {Phys. Rev. D}\ }\textbf {\bibinfo {volume} {98}},\ \bibinfo
  {pages} {114038} (\bibinfo {year} {2018})},\ \Eprint
  {https://arxiv.org/abs/1810.05971} {arXiv:1810.05971 [nucl-th]} \BibitemShut
  {NoStop}%
\bibitem [{\citenamefont {Davies}\ \emph {et~al.}(1996)\citenamefont {Davies},
  \citenamefont {Hornbostel}, \citenamefont {Lepage}, \citenamefont {Lidsey},
  \citenamefont {Shigemitsu},\ and\ \citenamefont {Sloan}}]{Davies:1996gi}%
  \BibitemOpen
  \bibfield  {author} {\bibinfo {author} {\bibfnamefont {C.~T.~H.}\
  \bibnamefont {Davies}}, \bibinfo {author} {\bibfnamefont {K.}~\bibnamefont
  {Hornbostel}}, \bibinfo {author} {\bibfnamefont {G.~P.}\ \bibnamefont
  {Lepage}}, \bibinfo {author} {\bibfnamefont {A.~J.}\ \bibnamefont {Lidsey}},
  \bibinfo {author} {\bibfnamefont {J.}~\bibnamefont {Shigemitsu}},\ and\
  \bibinfo {author} {\bibfnamefont {J.~H.}\ \bibnamefont {Sloan}},\ }\href
  {https://doi.org/10.1016/0370-2693(96)00650-8} {\bibfield  {journal}
  {\bibinfo  {journal} {Phys. Lett. B}\ }\textbf {\bibinfo {volume} {382}},\
  \bibinfo {pages} {131} (\bibinfo {year} {1996})},\ \Eprint
  {https://arxiv.org/abs/hep-lat/9602020} {arXiv:hep-lat/9602020} \BibitemShut
  {NoStop}%
\bibitem [{\citenamefont {Jones}\ and\ \citenamefont
  {Woloshyn}(1999)}]{Jones:1998ub}%
  \BibitemOpen
  \bibfield  {author} {\bibinfo {author} {\bibfnamefont {B.~D.}\ \bibnamefont
  {Jones}}\ and\ \bibinfo {author} {\bibfnamefont {R.~M.}\ \bibnamefont
  {Woloshyn}},\ }\href {https://doi.org/10.1103/PhysRevD.60.014502} {\bibfield
  {journal} {\bibinfo  {journal} {Phys. Rev. D}\ }\textbf {\bibinfo {volume}
  {60}},\ \bibinfo {pages} {014502} (\bibinfo {year} {1999})},\ \Eprint
  {https://arxiv.org/abs/hep-lat/9812008} {arXiv:hep-lat/9812008} \BibitemShut
  {NoStop}%
\bibitem [{\citenamefont {Gregory}\ \emph {et~al.}(2010)\citenamefont
  {Gregory}, \citenamefont {Davies}, \citenamefont {Follana}, \citenamefont
  {Gamiz}, \citenamefont {Kendall}, \citenamefont {Lepage}, \citenamefont {Na},
  \citenamefont {Shigemitsu},\ and\ \citenamefont {Wong}}]{Gregory:2009hq}%
  \BibitemOpen
  \bibfield  {author} {\bibinfo {author} {\bibfnamefont {E.~B.}\ \bibnamefont
  {Gregory}}, \bibinfo {author} {\bibfnamefont {C.~T.~H.}\ \bibnamefont
  {Davies}}, \bibinfo {author} {\bibfnamefont {E.}~\bibnamefont {Follana}},
  \bibinfo {author} {\bibfnamefont {E.}~\bibnamefont {Gamiz}}, \bibinfo
  {author} {\bibfnamefont {I.~D.}\ \bibnamefont {Kendall}}, \bibinfo {author}
  {\bibfnamefont {G.~P.}\ \bibnamefont {Lepage}}, \bibinfo {author}
  {\bibfnamefont {H.}~\bibnamefont {Na}}, \bibinfo {author} {\bibfnamefont
  {J.}~\bibnamefont {Shigemitsu}},\ and\ \bibinfo {author} {\bibfnamefont
  {K.~Y.}\ \bibnamefont {Wong}},\ }\href
  {https://doi.org/10.1103/PhysRevLett.104.022001} {\bibfield  {journal}
  {\bibinfo  {journal} {Phys. Rev. Lett.}\ }\textbf {\bibinfo {volume} {104}},\
  \bibinfo {pages} {022001} (\bibinfo {year} {2010})},\ \Eprint
  {https://arxiv.org/abs/0909.4462} {arXiv:0909.4462 [hep-lat]} \BibitemShut
  {NoStop}%
\bibitem [{\citenamefont {Mathur}\ \emph {et~al.}(2018)\citenamefont {Mathur},
  \citenamefont {Padmanath},\ and\ \citenamefont {Mondal}}]{Mathur:2018epb}%
  \BibitemOpen
  \bibfield  {author} {\bibinfo {author} {\bibfnamefont {N.}~\bibnamefont
  {Mathur}}, \bibinfo {author} {\bibfnamefont {M.}~\bibnamefont {Padmanath}},\
  and\ \bibinfo {author} {\bibfnamefont {S.}~\bibnamefont {Mondal}},\ }\href
  {https://doi.org/10.1103/PhysRevLett.121.202002} {\bibfield  {journal}
  {\bibinfo  {journal} {Phys. Rev. Lett.}\ }\textbf {\bibinfo {volume} {121}},\
  \bibinfo {pages} {202002} (\bibinfo {year} {2018})},\ \Eprint
  {https://arxiv.org/abs/1806.04151} {arXiv:1806.04151 [hep-lat]} \BibitemShut
  {NoStop}%
\bibitem [{\citenamefont {McNeile}\ \emph {et~al.}(2012)\citenamefont
  {McNeile}, \citenamefont {Davies}, \citenamefont {Follana}, \citenamefont
  {Hornbostel},\ and\ \citenamefont {Lepage}}]{McNeile:2012qf}%
  \BibitemOpen
  \bibfield  {author} {\bibinfo {author} {\bibfnamefont {C.}~\bibnamefont
  {McNeile}}, \bibinfo {author} {\bibfnamefont {C.~T.~H.}\ \bibnamefont
  {Davies}}, \bibinfo {author} {\bibfnamefont {E.}~\bibnamefont {Follana}},
  \bibinfo {author} {\bibfnamefont {K.}~\bibnamefont {Hornbostel}},\ and\
  \bibinfo {author} {\bibfnamefont {G.~P.}\ \bibnamefont {Lepage}} (\bibinfo
  {collaboration} {HPQCD}),\ }\href
  {https://doi.org/10.1103/PhysRevD.86.074503} {\bibfield  {journal} {\bibinfo
  {journal} {Phys. Rev. D}\ }\textbf {\bibinfo {volume} {86}},\ \bibinfo
  {pages} {074503} (\bibinfo {year} {2012})},\ \Eprint
  {https://arxiv.org/abs/1207.0994} {arXiv:1207.0994 [hep-lat]} \BibitemShut
  {NoStop}%
\bibitem [{\citenamefont {Colquhoun}\ \emph {et~al.}(2015)\citenamefont
  {Colquhoun}, \citenamefont {Davies}, \citenamefont {Dowdall}, \citenamefont
  {Kettle}, \citenamefont {Koponen}, \citenamefont {Lepage},\ and\
  \citenamefont {Lytle}}]{Colquhoun:2015oha}%
  \BibitemOpen
  \bibfield  {author} {\bibinfo {author} {\bibfnamefont {B.}~\bibnamefont
  {Colquhoun}}, \bibinfo {author} {\bibfnamefont {C.~T.~H.}\ \bibnamefont
  {Davies}}, \bibinfo {author} {\bibfnamefont {R.~J.}\ \bibnamefont {Dowdall}},
  \bibinfo {author} {\bibfnamefont {J.}~\bibnamefont {Kettle}}, \bibinfo
  {author} {\bibfnamefont {J.}~\bibnamefont {Koponen}}, \bibinfo {author}
  {\bibfnamefont {G.~P.}\ \bibnamefont {Lepage}},\ and\ \bibinfo {author}
  {\bibfnamefont {A.~T.}\ \bibnamefont {Lytle}} (\bibinfo {collaboration}
  {HPQCD}),\ }\href {https://doi.org/10.1103/PhysRevD.91.114509} {\bibfield
  {journal} {\bibinfo  {journal} {Phys. Rev. D}\ }\textbf {\bibinfo {volume}
  {91}},\ \bibinfo {pages} {114509} (\bibinfo {year} {2015})},\ \Eprint
  {https://arxiv.org/abs/1503.05762} {arXiv:1503.05762 [hep-lat]} \BibitemShut
  {NoStop}%
\bibitem [{\citenamefont {Colangelo}\ \emph {et~al.}(1993)\citenamefont
  {Colangelo}, \citenamefont {Nardulli},\ and\ \citenamefont
  {Paver}}]{Colangelo:1992cx}%
  \BibitemOpen
  \bibfield  {author} {\bibinfo {author} {\bibfnamefont {P.}~\bibnamefont
  {Colangelo}}, \bibinfo {author} {\bibfnamefont {G.}~\bibnamefont
  {Nardulli}},\ and\ \bibinfo {author} {\bibfnamefont {N.}~\bibnamefont
  {Paver}},\ }\href {https://doi.org/10.1007/BF01555737} {\bibfield  {journal}
  {\bibinfo  {journal} {Z. Phys. C}\ }\textbf {\bibinfo {volume} {57}},\
  \bibinfo {pages} {43} (\bibinfo {year} {1993})}\BibitemShut {NoStop}%
\bibitem [{\citenamefont {Chabab}(1994)}]{Chabab:1993nz}%
  \BibitemOpen
  \bibfield  {author} {\bibinfo {author} {\bibfnamefont {M.}~\bibnamefont
  {Chabab}},\ }\href {https://doi.org/10.1016/0370-2693(94)90093-0} {\bibfield
  {journal} {\bibinfo  {journal} {Phys. Lett. B}\ }\textbf {\bibinfo {volume}
  {325}},\ \bibinfo {pages} {205} (\bibinfo {year} {1994})}\BibitemShut
  {NoStop}%
\bibitem [{\citenamefont {Kiselev}\ and\ \citenamefont
  {Tkabladze}(1993)}]{Kiselev:1993ea}%
  \BibitemOpen
  \bibfield  {author} {\bibinfo {author} {\bibfnamefont {V.~V.}\ \bibnamefont
  {Kiselev}}\ and\ \bibinfo {author} {\bibfnamefont {A.~V.}\ \bibnamefont
  {Tkabladze}},\ }\href {https://doi.org/10.1103/PhysRevD.48.5208} {\bibfield
  {journal} {\bibinfo  {journal} {Phys. Rev. D}\ }\textbf {\bibinfo {volume}
  {48}},\ \bibinfo {pages} {5208} (\bibinfo {year} {1993})}\BibitemShut
  {NoStop}%
\bibitem [{\citenamefont {Bagan}\ \emph {et~al.}(1994)\citenamefont {Bagan},
  \citenamefont {Dosch}, \citenamefont {Gosdzinsky}, \citenamefont {Narison},\
  and\ \citenamefont {Richard}}]{Bagan:1994dy}%
  \BibitemOpen
  \bibfield  {author} {\bibinfo {author} {\bibfnamefont {E.}~\bibnamefont
  {Bagan}}, \bibinfo {author} {\bibfnamefont {H.~G.}\ \bibnamefont {Dosch}},
  \bibinfo {author} {\bibfnamefont {P.}~\bibnamefont {Gosdzinsky}}, \bibinfo
  {author} {\bibfnamefont {S.}~\bibnamefont {Narison}},\ and\ \bibinfo {author}
  {\bibfnamefont {J.~M.}\ \bibnamefont {Richard}},\ }\href
  {https://doi.org/10.1007/BF01557235} {\bibfield  {journal} {\bibinfo
  {journal} {Z. Phys. C}\ }\textbf {\bibinfo {volume} {64}},\ \bibinfo {pages}
  {57} (\bibinfo {year} {1994})},\ \Eprint
  {https://arxiv.org/abs/hep-ph/9403208} {arXiv:hep-ph/9403208} \BibitemShut
  {NoStop}%
\bibitem [{\citenamefont {Wang}(2013{\natexlab{a}})}]{Wang:2012kw}%
  \BibitemOpen
  \bibfield  {author} {\bibinfo {author} {\bibfnamefont {Z.-G.}\ \bibnamefont
  {Wang}},\ }\href {https://doi.org/10.1140/epja/i2013-13131-7} {\bibfield
  {journal} {\bibinfo  {journal} {Eur. Phys. J. A}\ }\textbf {\bibinfo {volume}
  {49}},\ \bibinfo {pages} {131} (\bibinfo {year} {2013}{\natexlab{a}})},\
  \Eprint {https://arxiv.org/abs/1203.6252} {arXiv:1203.6252 [hep-ph]}
  \BibitemShut {NoStop}%
\bibitem [{\citenamefont {Wang}(2013{\natexlab{b}})}]{Wang:2013cdy}%
  \BibitemOpen
  \bibfield  {author} {\bibinfo {author} {\bibfnamefont {Z.-G.}\ \bibnamefont
  {Wang}},\ }\href {https://doi.org/10.5506/APhysPolB.44.1971} {\bibfield
  {journal} {\bibinfo  {journal} {Acta Phys. Polon. B}\ }\textbf {\bibinfo
  {volume} {44}},\ \bibinfo {pages} {1971} (\bibinfo {year}
  {2013}{\natexlab{b}})},\ \Eprint {https://arxiv.org/abs/1303.4146}
  {arXiv:1303.4146 [hep-ph]} \BibitemShut {NoStop}%
\bibitem [{\citenamefont {Baker}\ \emph {et~al.}(2014)\citenamefont {Baker},
  \citenamefont {Bordes}, \citenamefont {Dominguez}, \citenamefont
  {Penarrocha},\ and\ \citenamefont {Schilcher}}]{Baker:2013mwa}%
  \BibitemOpen
  \bibfield  {author} {\bibinfo {author} {\bibfnamefont {M.~J.}\ \bibnamefont
  {Baker}}, \bibinfo {author} {\bibfnamefont {J.}~\bibnamefont {Bordes}},
  \bibinfo {author} {\bibfnamefont {C.~A.}\ \bibnamefont {Dominguez}}, \bibinfo
  {author} {\bibfnamefont {J.}~\bibnamefont {Penarrocha}},\ and\ \bibinfo
  {author} {\bibfnamefont {K.}~\bibnamefont {Schilcher}},\ }\href
  {https://doi.org/10.1007/JHEP07(2014)032} {\bibfield  {journal} {\bibinfo
  {journal} {JHEP}\ }\textbf {\bibinfo {volume} {07}},\ \bibinfo {pages}
  {032}},\ \Eprint {https://arxiv.org/abs/1310.0941} {arXiv:1310.0941 [hep-ph]}
  \BibitemShut {NoStop}%
\bibitem [{\citenamefont {Aliev}\ \emph {et~al.}(2019)\citenamefont {Aliev},
  \citenamefont {Barakat},\ and\ \citenamefont {Bilmis}}]{Aliev:2019wcm}%
  \BibitemOpen
  \bibfield  {author} {\bibinfo {author} {\bibfnamefont {T.~M.}\ \bibnamefont
  {Aliev}}, \bibinfo {author} {\bibfnamefont {T.}~\bibnamefont {Barakat}},\
  and\ \bibinfo {author} {\bibfnamefont {S.}~\bibnamefont {Bilmis}},\ }\href
  {https://doi.org/10.1016/j.nuclphysb.2019.114726} {\bibfield  {journal}
  {\bibinfo  {journal} {Nucl. Phys. B}\ }\textbf {\bibinfo {volume} {947}},\
  \bibinfo {pages} {114726} (\bibinfo {year} {2019})},\ \Eprint
  {https://arxiv.org/abs/1905.11750} {arXiv:1905.11750 [hep-ph]} \BibitemShut
  {NoStop}%
\bibitem [{\citenamefont {Narison}(2020)}]{Narison:2019tym}%
  \BibitemOpen
  \bibfield  {author} {\bibinfo {author} {\bibfnamefont {S.}~\bibnamefont
  {Narison}},\ }\href {https://doi.org/10.1016/j.physletb.2020.135221}
  {\bibfield  {journal} {\bibinfo  {journal} {Phys. Lett. B}\ }\textbf
  {\bibinfo {volume} {802}},\ \bibinfo {pages} {135221} (\bibinfo {year}
  {2020})},\ \Eprint {https://arxiv.org/abs/1906.03614} {arXiv:1906.03614
  [hep-ph]} \BibitemShut {NoStop}%
\bibitem [{\citenamefont {Wang}(2024)}]{Wang:2024fwc}%
  \BibitemOpen
  \bibfield  {author} {\bibinfo {author} {\bibfnamefont {Z.-G.}\ \bibnamefont
  {Wang}},\ }\href {https://doi.org/10.1088/1674-1137/ad5a71} {\bibfield
  {journal} {\bibinfo  {journal} {Chin. Phys. C}\ }\textbf {\bibinfo {volume}
  {48}},\ \bibinfo {pages} {103104} (\bibinfo {year} {2024})},\ \Eprint
  {https://arxiv.org/abs/2401.12571} {arXiv:2401.12571 [hep-ph]} \BibitemShut
  {NoStop}%
\bibitem [{\citenamefont {{\"O}zdem}(2025)}]{Ozdem:2024qaa}%
  \BibitemOpen
  \bibfield  {author} {\bibinfo {author} {\bibfnamefont {U.}~\bibnamefont
  {{\"O}zdem}},\ }\href {https://doi.org/10.1140/epjc/s10052-025-13959-8}
  {\bibfield  {journal} {\bibinfo  {journal} {Eur. Phys. J. C}\ }\textbf
  {\bibinfo {volume} {85}},\ \bibinfo {pages} {245} (\bibinfo {year} {2025})},\
  \Eprint {https://arxiv.org/abs/2411.06123} {arXiv:2411.06123 [hep-ph]}
  \BibitemShut {NoStop}%
\bibitem [{\citenamefont {Brambilla}\ and\ \citenamefont
  {Vairo}(2000)}]{Brambilla:2000db}%
  \BibitemOpen
  \bibfield  {author} {\bibinfo {author} {\bibfnamefont {N.}~\bibnamefont
  {Brambilla}}\ and\ \bibinfo {author} {\bibfnamefont {A.}~\bibnamefont
  {Vairo}},\ }\href {https://doi.org/10.1103/PhysRevD.62.094019} {\bibfield
  {journal} {\bibinfo  {journal} {Phys. Rev. D}\ }\textbf {\bibinfo {volume}
  {62}},\ \bibinfo {pages} {094019} (\bibinfo {year} {2000})},\ \Eprint
  {https://arxiv.org/abs/hep-ph/0002075} {arXiv:hep-ph/0002075} \BibitemShut
  {NoStop}%
\bibitem [{\citenamefont {Penin}\ \emph {et~al.}(2004)\citenamefont {Penin},
  \citenamefont {Pineda}, \citenamefont {Smirnov},\ and\ \citenamefont
  {Steinhauser}}]{Penin:2004xi}%
  \BibitemOpen
  \bibfield  {author} {\bibinfo {author} {\bibfnamefont {A.~A.}\ \bibnamefont
  {Penin}}, \bibinfo {author} {\bibfnamefont {A.}~\bibnamefont {Pineda}},
  \bibinfo {author} {\bibfnamefont {V.~A.}\ \bibnamefont {Smirnov}},\ and\
  \bibinfo {author} {\bibfnamefont {M.}~\bibnamefont {Steinhauser}},\ }\href
  {https://doi.org/10.1016/j.physletb.2004.04.066} {\bibfield  {journal}
  {\bibinfo  {journal} {Phys. Lett. B}\ }\textbf {\bibinfo {volume} {593}},\
  \bibinfo {pages} {124} (\bibinfo {year} {2004})},\ \bibinfo {note} {[Erratum:
  Phys.Lett.B 677, 343 (2009)]},\ \Eprint
  {https://arxiv.org/abs/hep-ph/0403080} {arXiv:hep-ph/0403080} \BibitemShut
  {NoStop}%
\bibitem [{\citenamefont {Onishchenko}\ and\ \citenamefont
  {Veretin}(2007)}]{Onishchenko:2003ui}%
  \BibitemOpen
  \bibfield  {author} {\bibinfo {author} {\bibfnamefont {A.~I.}\ \bibnamefont
  {Onishchenko}}\ and\ \bibinfo {author} {\bibfnamefont {O.~L.}\ \bibnamefont
  {Veretin}},\ }\href {https://doi.org/10.1140/epjc/s10052-007-0255-1}
  {\bibfield  {journal} {\bibinfo  {journal} {Eur. Phys. J. C}\ }\textbf
  {\bibinfo {volume} {50}},\ \bibinfo {pages} {801} (\bibinfo {year} {2007})},\
  \Eprint {https://arxiv.org/abs/hep-ph/0302132} {arXiv:hep-ph/0302132}
  \BibitemShut {NoStop}%
\bibitem [{\citenamefont {Lee}\ \emph {et~al.}(2011)\citenamefont {Lee},
  \citenamefont {Sang},\ and\ \citenamefont {Kim}}]{Lee:2010ts}%
  \BibitemOpen
  \bibfield  {author} {\bibinfo {author} {\bibfnamefont {J.}~\bibnamefont
  {Lee}}, \bibinfo {author} {\bibfnamefont {W.}~\bibnamefont {Sang}},\ and\
  \bibinfo {author} {\bibfnamefont {S.}~\bibnamefont {Kim}},\ }\href
  {https://doi.org/10.1007/JHEP01(2011)113} {\bibfield  {journal} {\bibinfo
  {journal} {JHEP}\ }\textbf {\bibinfo {volume} {01}},\ \bibinfo {pages}
  {113}},\ \Eprint {https://arxiv.org/abs/1011.2274} {arXiv:1011.2274 [hep-ph]}
  \BibitemShut {NoStop}%
\bibitem [{\citenamefont {Chen}\ and\ \citenamefont
  {Qiao}(2015)}]{Chen:2015csa}%
  \BibitemOpen
  \bibfield  {author} {\bibinfo {author} {\bibfnamefont {L.-B.}\ \bibnamefont
  {Chen}}\ and\ \bibinfo {author} {\bibfnamefont {C.-F.}\ \bibnamefont
  {Qiao}},\ }\href {https://doi.org/10.1016/j.physletb.2015.07.043} {\bibfield
  {journal} {\bibinfo  {journal} {Phys. Lett. B}\ }\textbf {\bibinfo {volume}
  {748}},\ \bibinfo {pages} {443} (\bibinfo {year} {2015})},\ \Eprint
  {https://arxiv.org/abs/1503.05122} {arXiv:1503.05122 [hep-ph]} \BibitemShut
  {NoStop}%
\bibitem [{\citenamefont {Tao}\ \emph {et~al.}(2022)\citenamefont {Tao},
  \citenamefont {Zhu},\ and\ \citenamefont {Xiao}}]{Tao:2022qxa}%
  \BibitemOpen
  \bibfield  {author} {\bibinfo {author} {\bibfnamefont {W.}~\bibnamefont
  {Tao}}, \bibinfo {author} {\bibfnamefont {R.}~\bibnamefont {Zhu}},\ and\
  \bibinfo {author} {\bibfnamefont {Z.-J.}\ \bibnamefont {Xiao}},\ }\href
  {https://doi.org/10.1103/PhysRevD.106.114037} {\bibfield  {journal} {\bibinfo
   {journal} {Phys. Rev. D}\ }\textbf {\bibinfo {volume} {106}},\ \bibinfo
  {pages} {114037} (\bibinfo {year} {2022})},\ \Eprint
  {https://arxiv.org/abs/2209.15521} {arXiv:2209.15521 [hep-ph]} \BibitemShut
  {NoStop}%
\bibitem [{\citenamefont {Tao}\ \emph {et~al.}(2023)\citenamefont {Tao},
  \citenamefont {Zhu},\ and\ \citenamefont {Xiao}}]{Tao:2022hos}%
  \BibitemOpen
  \bibfield  {author} {\bibinfo {author} {\bibfnamefont {W.}~\bibnamefont
  {Tao}}, \bibinfo {author} {\bibfnamefont {R.}~\bibnamefont {Zhu}},\ and\
  \bibinfo {author} {\bibfnamefont {Z.-J.}\ \bibnamefont {Xiao}},\ }\href
  {https://doi.org/10.1140/epjc/s10052-023-11442-w} {\bibfield  {journal}
  {\bibinfo  {journal} {Eur. Phys. J. C}\ }\textbf {\bibinfo {volume} {83}},\
  \bibinfo {pages} {294} (\bibinfo {year} {2023})},\ \Eprint
  {https://arxiv.org/abs/2301.00220} {arXiv:2301.00220 [hep-ph]} \BibitemShut
  {NoStop}%
\bibitem [{\citenamefont {Sang}\ \emph {et~al.}(2023)\citenamefont {Sang},
  \citenamefont {Zhang},\ and\ \citenamefont {Zhou}}]{Sang:2022tnh}%
  \BibitemOpen
  \bibfield  {author} {\bibinfo {author} {\bibfnamefont {W.-L.}\ \bibnamefont
  {Sang}}, \bibinfo {author} {\bibfnamefont {H.-F.}\ \bibnamefont {Zhang}},\
  and\ \bibinfo {author} {\bibfnamefont {M.-Z.}\ \bibnamefont {Zhou}},\ }\href
  {https://doi.org/10.1016/j.physletb.2023.137812} {\bibfield  {journal}
  {\bibinfo  {journal} {Phys. Lett. B}\ }\textbf {\bibinfo {volume} {839}},\
  \bibinfo {pages} {137812} (\bibinfo {year} {2023})},\ \Eprint
  {https://arxiv.org/abs/2210.02979} {arXiv:2210.02979 [hep-ph]} \BibitemShut
  {NoStop}%
\bibitem [{\citenamefont {Feng}\ \emph {et~al.}(2022)\citenamefont {Feng},
  \citenamefont {Jia}, \citenamefont {Mo}, \citenamefont {Pan}, \citenamefont
  {Sang},\ and\ \citenamefont {Zhang}}]{Feng:2022ruy}%
  \BibitemOpen
  \bibfield  {author} {\bibinfo {author} {\bibfnamefont {F.}~\bibnamefont
  {Feng}}, \bibinfo {author} {\bibfnamefont {Y.}~\bibnamefont {Jia}}, \bibinfo
  {author} {\bibfnamefont {Z.}~\bibnamefont {Mo}}, \bibinfo {author}
  {\bibfnamefont {J.}~\bibnamefont {Pan}}, \bibinfo {author} {\bibfnamefont
  {W.-L.}\ \bibnamefont {Sang}},\ and\ \bibinfo {author} {\bibfnamefont
  {J.-Y.}\ \bibnamefont {Zhang}},\ }\href@noop {} {\  (\bibinfo {year}
  {2022})},\ \Eprint {https://arxiv.org/abs/2208.04302} {arXiv:2208.04302
  [hep-ph]} \BibitemShut {NoStop}%
\bibitem [{\citenamefont {Tao}\ and\ \citenamefont {Xiao}(2024)}]{Tao:2023pzv}%
  \BibitemOpen
  \bibfield  {author} {\bibinfo {author} {\bibfnamefont {W.}~\bibnamefont
  {Tao}}\ and\ \bibinfo {author} {\bibfnamefont {Z.-J.}\ \bibnamefont {Xiao}},\
  }\href {https://doi.org/10.1007/JHEP06(2024)012} {\bibfield  {journal}
  {\bibinfo  {journal} {JHEP}\ }\textbf {\bibinfo {volume} {06}},\ \bibinfo
  {pages} {012}},\ \Eprint {https://arxiv.org/abs/2310.17500} {arXiv:2310.17500
  [hep-ph]} \BibitemShut {NoStop}%
\bibitem [{\citenamefont {Ikhdair}\ and\ \citenamefont
  {Sever}(2004)}]{Ikhdair:2003ry}%
  \BibitemOpen
  \bibfield  {author} {\bibinfo {author} {\bibfnamefont {S.~M.}\ \bibnamefont
  {Ikhdair}}\ and\ \bibinfo {author} {\bibfnamefont {R.}~\bibnamefont
  {Sever}},\ }\href {https://doi.org/10.1142/S0217751X0401780X} {\bibfield
  {journal} {\bibinfo  {journal} {Int. J. Mod. Phys. A}\ }\textbf {\bibinfo
  {volume} {19}},\ \bibinfo {pages} {1771} (\bibinfo {year} {2004})},\ \Eprint
  {https://arxiv.org/abs/hep-ph/0310295} {arXiv:hep-ph/0310295} \BibitemShut
  {NoStop}%
\bibitem [{\citenamefont {Ikhdair}\ and\ \citenamefont
  {Sever}(2006)}]{Ikhdair:2006nx}%
  \BibitemOpen
  \bibfield  {author} {\bibinfo {author} {\bibfnamefont {S.~M.}\ \bibnamefont
  {Ikhdair}}\ and\ \bibinfo {author} {\bibfnamefont {R.}~\bibnamefont
  {Sever}},\ }\href {https://doi.org/10.1142/S0217751X06034100} {\bibfield
  {journal} {\bibinfo  {journal} {Int. J. Mod. Phys. A}\ }\textbf {\bibinfo
  {volume} {21}},\ \bibinfo {pages} {6699} (\bibinfo {year} {2006})},\ \Eprint
  {https://arxiv.org/abs/hep-ph/0702166} {arXiv:hep-ph/0702166} \BibitemShut
  {NoStop}%
\bibitem [{\citenamefont {Abd El-Hady}\ \emph {et~al.}(1999)\citenamefont {Abd
  El-Hady}, \citenamefont {Lodhi},\ and\ \citenamefont
  {Vary}}]{AbdEl-Hady:1998uiq}%
  \BibitemOpen
  \bibfield  {author} {\bibinfo {author} {\bibfnamefont {A.}~\bibnamefont {Abd
  El-Hady}}, \bibinfo {author} {\bibfnamefont {M.~A.~K.}\ \bibnamefont
  {Lodhi}},\ and\ \bibinfo {author} {\bibfnamefont {J.~P.}\ \bibnamefont
  {Vary}},\ }\href {https://doi.org/10.1103/PhysRevD.59.094001} {\bibfield
  {journal} {\bibinfo  {journal} {Phys. Rev. D}\ }\textbf {\bibinfo {volume}
  {59}},\ \bibinfo {pages} {094001} (\bibinfo {year} {1999})},\ \Eprint
  {https://arxiv.org/abs/hep-ph/9807225} {arXiv:hep-ph/9807225} \BibitemShut
  {NoStop}%
\bibitem [{\citenamefont {Wang}(2007)}]{Wang:2007av}%
  \BibitemOpen
  \bibfield  {author} {\bibinfo {author} {\bibfnamefont {G.-L.}\ \bibnamefont
  {Wang}},\ }\href {https://doi.org/10.1016/j.physletb.2007.05.001} {\bibfield
  {journal} {\bibinfo  {journal} {Phys. Lett. B}\ }\textbf {\bibinfo {volume}
  {650}},\ \bibinfo {pages} {15} (\bibinfo {year} {2007})},\ \Eprint
  {https://arxiv.org/abs/0705.2621} {arXiv:0705.2621 [hep-ph]} \BibitemShut
  {NoStop}%
\bibitem [{\citenamefont {Wang}\ \emph {et~al.}(2022)\citenamefont {Wang},
  \citenamefont {Wang}, \citenamefont {Li},\ and\ \citenamefont
  {Chang}}]{Wang:2022cxy}%
  \BibitemOpen
  \bibfield  {author} {\bibinfo {author} {\bibfnamefont {G.-L.}\ \bibnamefont
  {Wang}}, \bibinfo {author} {\bibfnamefont {T.}~\bibnamefont {Wang}}, \bibinfo
  {author} {\bibfnamefont {Q.}~\bibnamefont {Li}},\ and\ \bibinfo {author}
  {\bibfnamefont {C.-H.}\ \bibnamefont {Chang}},\ }\href
  {https://doi.org/10.1007/JHEP05(2022)006} {\bibfield  {journal} {\bibinfo
  {journal} {JHEP}\ }\textbf {\bibinfo {volume} {05}},\ \bibinfo {pages}
  {006}},\ \Eprint {https://arxiv.org/abs/2201.02318} {arXiv:2201.02318
  [hep-ph]} \BibitemShut {NoStop}%
\bibitem [{\citenamefont {Badalian}\ \emph {et~al.}(2007)\citenamefont
  {Badalian}, \citenamefont {Bakker},\ and\ \citenamefont
  {Simonov}}]{Badalian:2007km}%
  \BibitemOpen
  \bibfield  {author} {\bibinfo {author} {\bibfnamefont {A.~M.}\ \bibnamefont
  {Badalian}}, \bibinfo {author} {\bibfnamefont {B.~L.~G.}\ \bibnamefont
  {Bakker}},\ and\ \bibinfo {author} {\bibfnamefont {Y.~A.}\ \bibnamefont
  {Simonov}},\ }\href {https://doi.org/10.1103/PhysRevD.75.116001} {\bibfield
  {journal} {\bibinfo  {journal} {Phys. Rev. D}\ }\textbf {\bibinfo {volume}
  {75}},\ \bibinfo {pages} {116001} (\bibinfo {year} {2007})},\ \Eprint
  {https://arxiv.org/abs/hep-ph/0702157} {arXiv:hep-ph/0702157} \BibitemShut
  {NoStop}%
\bibitem [{\citenamefont {Lodha}\ \emph {et~al.}(2026)\citenamefont {Lodha},
  \citenamefont {Parmar},\ and\ \citenamefont {Rai}}]{Lodha:2026epjp}%
  \BibitemOpen
  \bibfield  {author} {\bibinfo {author} {\bibfnamefont {C.}~\bibnamefont
  {Lodha}}, \bibinfo {author} {\bibfnamefont {M.}~\bibnamefont {Parmar}},\ and\
  \bibinfo {author} {\bibfnamefont {A.~K.}\ \bibnamefont {Rai}},\ }\href
  {https://doi.org/10.1140/epjp/s13360-026-07724-3} {\bibfield  {journal}
  {\bibinfo  {journal} {Eur. Phys. J. Plus}\ }\textbf {\bibinfo {volume}
  {141}},\ \bibinfo {pages} {453} (\bibinfo {year} {2026})}\BibitemShut
  {NoStop}%
\bibitem [{\citenamefont {Chaturvedi}\ and\ \citenamefont
  {Rai}(2022)}]{Chaturvedi:2022pmn}%
  \BibitemOpen
  \bibfield  {author} {\bibinfo {author} {\bibfnamefont {R.}~\bibnamefont
  {Chaturvedi}}\ and\ \bibinfo {author} {\bibfnamefont {A.~K.}\ \bibnamefont
  {Rai}},\ }\href {https://doi.org/10.1140/epja/s10050-022-00884-7} {\bibfield
  {journal} {\bibinfo  {journal} {Eur. Phys. J. A}\ }\textbf {\bibinfo {volume}
  {58}},\ \bibinfo {pages} {228} (\bibinfo {year} {2022})},\ \Eprint
  {https://arxiv.org/abs/2211.04099} {arXiv:2211.04099 [hep-ph]} \BibitemShut
  {NoStop}%
\bibitem [{\citenamefont {Bokade}\ and\ \citenamefont
  {Bhaghyesh}(2025)}]{Bokade:2025lmn}%
  \BibitemOpen
  \bibfield  {author} {\bibinfo {author} {\bibfnamefont {C.~A.}\ \bibnamefont
  {Bokade}}\ and\ \bibinfo {author} {\bibnamefont {Bhaghyesh}},\ }\href
  {https://doi.org/10.1016/j.nuclphysa.2025.123109} {\bibfield  {journal}
  {\bibinfo  {journal} {Nucl. Phys. A}\ }\textbf {\bibinfo {volume} {1060}},\
  \bibinfo {pages} {123109} (\bibinfo {year} {2025})}\BibitemShut {NoStop}%
\bibitem [{\citenamefont {Li}\ \emph {et~al.}(2023{\natexlab{a}})\citenamefont
  {Li}, \citenamefont {Tang}, \citenamefont {Fang}, \citenamefont {Wang},
  \citenamefont {Pang},\ and\ \citenamefont {Liu}}]{Li:2022bre}%
  \BibitemOpen
  \bibfield  {author} {\bibinfo {author} {\bibfnamefont {T.-y.}\ \bibnamefont
  {Li}}, \bibinfo {author} {\bibfnamefont {L.}~\bibnamefont {Tang}}, \bibinfo
  {author} {\bibfnamefont {Z.-y.}\ \bibnamefont {Fang}}, \bibinfo {author}
  {\bibfnamefont {C.-h.}\ \bibnamefont {Wang}}, \bibinfo {author}
  {\bibfnamefont {C.-q.}\ \bibnamefont {Pang}},\ and\ \bibinfo {author}
  {\bibfnamefont {X.}~\bibnamefont {Liu}},\ }\href
  {https://doi.org/10.1103/PhysRevD.108.034019} {\bibfield  {journal} {\bibinfo
   {journal} {Phys. Rev. D}\ }\textbf {\bibinfo {volume} {108}},\ \bibinfo
  {pages} {034019} (\bibinfo {year} {2023}{\natexlab{a}})},\ \Eprint
  {https://arxiv.org/abs/2204.14258} {arXiv:2204.14258 [hep-ph]} \BibitemShut
  {NoStop}%
\bibitem [{\citenamefont {Asghar}\ \emph {et~al.}(2019)\citenamefont {Asghar},
  \citenamefont {Akram}, \citenamefont {Masud},\ and\ \citenamefont
  {Sultan}}]{Asghar:2019qjl}%
  \BibitemOpen
  \bibfield  {author} {\bibinfo {author} {\bibfnamefont {I.}~\bibnamefont
  {Asghar}}, \bibinfo {author} {\bibfnamefont {F.}~\bibnamefont {Akram}},
  \bibinfo {author} {\bibfnamefont {B.}~\bibnamefont {Masud}},\ and\ \bibinfo
  {author} {\bibfnamefont {M.~A.}\ \bibnamefont {Sultan}},\ }\href
  {https://doi.org/10.1103/PhysRevD.100.096002} {\bibfield  {journal} {\bibinfo
   {journal} {Phys. Rev. D}\ }\textbf {\bibinfo {volume} {100}},\ \bibinfo
  {pages} {096002} (\bibinfo {year} {2019})},\ \Eprint
  {https://arxiv.org/abs/1910.02680} {arXiv:1910.02680 [hep-ph]} \BibitemShut
  {NoStop}%
\bibitem [{\citenamefont {{\c{C}}ak{\i}r}\ and\ \citenamefont
  {Mutuk}(2026)}]{Cakir:2026fzd}%
  \BibitemOpen
  \bibfield  {author} {\bibinfo {author} {\bibfnamefont {{\"O}.}~\bibnamefont
  {{\c{C}}ak{\i}r}}\ and\ \bibinfo {author} {\bibfnamefont {H.}~\bibnamefont
  {Mutuk}},\ }\href {https://doi.org/10.1140/epjc/s10052-026-15944-1}
  {\bibfield  {journal} {\bibinfo  {journal} {Eur. Phys. J. C}\ }\textbf
  {\bibinfo {volume} {86}},\ \bibinfo {pages} {678} (\bibinfo {year} {2026})},\
  \Eprint {https://arxiv.org/abs/2604.18796} {arXiv:2604.18796 [hep-ph]}
  \BibitemShut {NoStop}%
\bibitem [{\citenamefont {Eichten}\ \emph {et~al.}(1978)\citenamefont
  {Eichten}, \citenamefont {Gottfried}, \citenamefont {Kinoshita},
  \citenamefont {Lane},\ and\ \citenamefont {Yan}}]{Eichten:1978tg}%
  \BibitemOpen
  \bibfield  {author} {\bibinfo {author} {\bibfnamefont {E.}~\bibnamefont
  {Eichten}}, \bibinfo {author} {\bibfnamefont {K.}~\bibnamefont {Gottfried}},
  \bibinfo {author} {\bibfnamefont {T.}~\bibnamefont {Kinoshita}}, \bibinfo
  {author} {\bibfnamefont {K.~D.}\ \bibnamefont {Lane}},\ and\ \bibinfo
  {author} {\bibfnamefont {T.-M.}\ \bibnamefont {Yan}},\ }\href
  {https://doi.org/10.1103/PhysRevD.17.3090} {\bibfield  {journal} {\bibinfo
  {journal} {Phys. Rev. D}\ }\textbf {\bibinfo {volume} {17}},\ \bibinfo
  {pages} {3090} (\bibinfo {year} {1978})},\ \bibinfo {note} {[Erratum:
  Phys.Rev.D 21, 313 (1980)]}\BibitemShut {NoStop}%
\bibitem [{\citenamefont {Buchmuller}\ and\ \citenamefont
  {Tye}(1981)}]{Buchmuller:1980su}%
  \BibitemOpen
  \bibfield  {author} {\bibinfo {author} {\bibfnamefont {W.}~\bibnamefont
  {Buchmuller}}\ and\ \bibinfo {author} {\bibfnamefont {S.~H.~H.}\ \bibnamefont
  {Tye}},\ }\href {https://doi.org/10.1103/PhysRevD.24.132} {\bibfield
  {journal} {\bibinfo  {journal} {Phys. Rev. D}\ }\textbf {\bibinfo {volume}
  {24}},\ \bibinfo {pages} {132} (\bibinfo {year} {1981})}\BibitemShut
  {NoStop}%
\bibitem [{\citenamefont {Bali}(2001)}]{Bali:2000gf}%
  \BibitemOpen
  \bibfield  {author} {\bibinfo {author} {\bibfnamefont {G.~S.}\ \bibnamefont
  {Bali}},\ }\href {https://doi.org/10.1016/S0370-1573(00)00079-X} {\bibfield
  {journal} {\bibinfo  {journal} {Phys. Rept.}\ }\textbf {\bibinfo {volume}
  {343}},\ \bibinfo {pages} {1} (\bibinfo {year} {2001})},\ \Eprint
  {https://arxiv.org/abs/hep-ph/0001312} {arXiv:hep-ph/0001312} \BibitemShut
  {NoStop}%
\bibitem [{\citenamefont {Barnes}\ \emph {et~al.}(2005)\citenamefont {Barnes},
  \citenamefont {Godfrey},\ and\ \citenamefont {Swanson}}]{Barnes:2005pb}%
  \BibitemOpen
  \bibfield  {author} {\bibinfo {author} {\bibfnamefont {T.}~\bibnamefont
  {Barnes}}, \bibinfo {author} {\bibfnamefont {S.}~\bibnamefont {Godfrey}},\
  and\ \bibinfo {author} {\bibfnamefont {E.~S.}\ \bibnamefont {Swanson}},\
  }\href {https://doi.org/10.1103/PhysRevD.72.054026} {\bibfield  {journal}
  {\bibinfo  {journal} {Phys. Rev. D}\ }\textbf {\bibinfo {volume} {72}},\
  \bibinfo {pages} {054026} (\bibinfo {year} {2005})},\ \Eprint
  {https://arxiv.org/abs/hep-ph/0505002} {arXiv:hep-ph/0505002} \BibitemShut
  {NoStop}%
\bibitem [{\citenamefont {Lucha}\ \emph {et~al.}(1991)\citenamefont {Lucha},
  \citenamefont {Schoberl},\ and\ \citenamefont {Gromes}}]{Lucha:1991vn}%
  \BibitemOpen
  \bibfield  {author} {\bibinfo {author} {\bibfnamefont {W.}~\bibnamefont
  {Lucha}}, \bibinfo {author} {\bibfnamefont {F.~F.}\ \bibnamefont
  {Schoberl}},\ and\ \bibinfo {author} {\bibfnamefont {D.}~\bibnamefont
  {Gromes}},\ }\href {https://doi.org/10.1016/0370-1573(91)90001-3} {\bibfield
  {journal} {\bibinfo  {journal} {Phys. Rept.}\ }\textbf {\bibinfo {volume}
  {200}},\ \bibinfo {pages} {127} (\bibinfo {year} {1991})}\BibitemShut
  {NoStop}%
\bibitem [{\citenamefont {Debastiani}\ and\ \citenamefont
  {Navarra}(2019)}]{Debastiani:2017msn}%
  \BibitemOpen
  \bibfield  {author} {\bibinfo {author} {\bibfnamefont {V.~R.}\ \bibnamefont
  {Debastiani}}\ and\ \bibinfo {author} {\bibfnamefont {F.~S.}\ \bibnamefont
  {Navarra}},\ }\href {https://doi.org/10.1088/1674-1137/43/1/013105}
  {\bibfield  {journal} {\bibinfo  {journal} {Chin. Phys. C}\ }\textbf
  {\bibinfo {volume} {43}},\ \bibinfo {pages} {013105} (\bibinfo {year}
  {2019})},\ \Eprint {https://arxiv.org/abs/1706.07553} {arXiv:1706.07553
  [hep-ph]} \BibitemShut {NoStop}%
\bibitem [{\citenamefont {Van~Royen}\ and\ \citenamefont
  {Weisskopf}(1967)}]{VanRoyen:1967nq}%
  \BibitemOpen
  \bibfield  {author} {\bibinfo {author} {\bibfnamefont {R.}~\bibnamefont
  {Van~Royen}}\ and\ \bibinfo {author} {\bibfnamefont {V.~F.}\ \bibnamefont
  {Weisskopf}},\ }\href {https://doi.org/10.1007/BF02823542} {\bibfield
  {journal} {\bibinfo  {journal} {Nuovo Cim. A}\ }\textbf {\bibinfo {volume}
  {50}},\ \bibinfo {pages} {617} (\bibinfo {year} {1967})},\ \bibinfo {note}
  {[Erratum: Nuovo Cim.A 51, 583 (1967)]}\BibitemShut {NoStop}%
\bibitem [{\citenamefont {Braaten}\ and\ \citenamefont
  {Fleming}(1995)}]{Braaten:1995ej}%
  \BibitemOpen
  \bibfield  {author} {\bibinfo {author} {\bibfnamefont {E.}~\bibnamefont
  {Braaten}}\ and\ \bibinfo {author} {\bibfnamefont {S.}~\bibnamefont
  {Fleming}},\ }\href {https://doi.org/10.1103/PhysRevD.52.181} {\bibfield
  {journal} {\bibinfo  {journal} {Phys. Rev. D}\ }\textbf {\bibinfo {volume}
  {52}},\ \bibinfo {pages} {181} (\bibinfo {year} {1995})},\ \Eprint
  {https://arxiv.org/abs/hep-ph/9501296} {arXiv:hep-ph/9501296} \BibitemShut
  {NoStop}%
\bibitem [{\citenamefont {Gershtein}\ \emph
  {et~al.}(1995{\natexlab{b}})\citenamefont {Gershtein}, \citenamefont
  {Kiselev}, \citenamefont {Likhoded},\ and\ \citenamefont
  {Tkabladze}}]{Gershtein:1994jw}%
  \BibitemOpen
  \bibfield  {author} {\bibinfo {author} {\bibfnamefont {S.~S.}\ \bibnamefont
  {Gershtein}}, \bibinfo {author} {\bibfnamefont {V.~V.}\ \bibnamefont
  {Kiselev}}, \bibinfo {author} {\bibfnamefont {A.~K.}\ \bibnamefont
  {Likhoded}},\ and\ \bibinfo {author} {\bibfnamefont {A.~V.}\ \bibnamefont
  {Tkabladze}},\ }\href {https://doi.org/10.1070/PU1995v038n01ABEH000063}
  {\bibfield  {journal} {\bibinfo  {journal} {Phys. Usp.}\ }\textbf {\bibinfo
  {volume} {38}},\ \bibinfo {pages} {1} (\bibinfo {year}
  {1995}{\natexlab{b}})},\ \Eprint {https://arxiv.org/abs/hep-ph/9504319}
  {arXiv:hep-ph/9504319} \BibitemShut {NoStop}%
\bibitem [{\citenamefont {Li}\ \emph {et~al.}(2023{\natexlab{b}})\citenamefont
  {Li}, \citenamefont {Li}, \citenamefont {Wang},\ and\ \citenamefont
  {Liu}}]{Li:2023wgq}%
  \BibitemOpen
  \bibfield  {author} {\bibinfo {author} {\bibfnamefont {X.-J.}\ \bibnamefont
  {Li}}, \bibinfo {author} {\bibfnamefont {Y.-S.}\ \bibnamefont {Li}}, \bibinfo
  {author} {\bibfnamefont {F.-L.}\ \bibnamefont {Wang}},\ and\ \bibinfo
  {author} {\bibfnamefont {X.}~\bibnamefont {Liu}},\ }\href
  {https://doi.org/10.1140/epjc/s10052-023-12237-9} {\bibfield  {journal}
  {\bibinfo  {journal} {Eur. Phys. J. C}\ }\textbf {\bibinfo {volume} {83}},\
  \bibinfo {pages} {1080} (\bibinfo {year} {2023}{\natexlab{b}})},\ \Eprint
  {https://arxiv.org/abs/2308.07206} {arXiv:2308.07206 [hep-ph]} \BibitemShut
  {NoStop}%
\bibitem [{\citenamefont {Sun}\ \emph {et~al.}(2023)\citenamefont {Sun},
  \citenamefont {Ni},\ and\ \citenamefont {Chen}}]{2209.06724}%
  \BibitemOpen
  \bibfield  {author} {\bibinfo {author} {\bibfnamefont {C.}~\bibnamefont
  {Sun}}, \bibinfo {author} {\bibfnamefont {R.-H.}\ \bibnamefont {Ni}},\ and\
  \bibinfo {author} {\bibfnamefont {M.}~\bibnamefont {Chen}},\ }\href
  {https://doi.org/10.1088/1674-1137/ac9dea} {\bibfield  {journal} {\bibinfo
  {journal} {Chin. Phys. C}\ }\textbf {\bibinfo {volume} {47}},\ \bibinfo
  {pages} {023101} (\bibinfo {year} {2023})},\ \Eprint
  {https://arxiv.org/abs/2209.06724} {arXiv:2209.06724 [hep-ph]} \BibitemShut
  {NoStop}%
\bibitem [{\citenamefont {Akbar}(2020)}]{2003.08491}%
  \BibitemOpen
  \bibfield  {author} {\bibinfo {author} {\bibfnamefont {N.}~\bibnamefont
  {Akbar}},\ }\href@noop {} {\  (\bibinfo {year} {2020})},\ \Eprint
  {https://arxiv.org/abs/2003.08491} {arXiv:2003.08491 [hep-ph]} \BibitemShut
  {NoStop}%
\bibitem [{\citenamefont {Patel}\ \emph {et~al.}(2009)\citenamefont {Patel},
  \citenamefont {Majethiya},\ and\ \citenamefont {Vinodkumar}}]{Patel:2008mv}%
  \BibitemOpen
  \bibfield  {author} {\bibinfo {author} {\bibfnamefont {B.}~\bibnamefont
  {Patel}}, \bibinfo {author} {\bibfnamefont {A.}~\bibnamefont {Majethiya}},\
  and\ \bibinfo {author} {\bibfnamefont {P.~C.}\ \bibnamefont {Vinodkumar}},\
  }\href {https://doi.org/10.1007/s12043-009-0061-4} {\bibfield  {journal}
  {\bibinfo  {journal} {Pramana}\ }\textbf {\bibinfo {volume} {72}},\ \bibinfo
  {pages} {679} (\bibinfo {year} {2009})},\ \Eprint
  {https://arxiv.org/abs/0808.2880} {arXiv:0808.2880 [hep-ph]} \BibitemShut
  {NoStop}%
\bibitem [{\citenamefont {Akbar}\ \emph {et~al.}(2019)\citenamefont {Akbar},
  \citenamefont {Akram}, \citenamefont {Masud},\ and\ \citenamefont
  {Atif~Sultan}}]{Akbar:2018hiw}%
  \BibitemOpen
  \bibfield  {author} {\bibinfo {author} {\bibfnamefont {N.}~\bibnamefont
  {Akbar}}, \bibinfo {author} {\bibfnamefont {F.}~\bibnamefont {Akram}},
  \bibinfo {author} {\bibfnamefont {B.}~\bibnamefont {Masud}},\ and\ \bibinfo
  {author} {\bibfnamefont {M.}~\bibnamefont {Atif~Sultan}},\ }\href
  {https://doi.org/10.1140/epja/i2019-12735-1} {\bibfield  {journal} {\bibinfo
  {journal} {Eur. Phys. J. A}\ }\textbf {\bibinfo {volume} {55}},\ \bibinfo
  {pages} {82} (\bibinfo {year} {2019})},\ \Eprint
  {https://arxiv.org/abs/1811.07552} {arXiv:1811.07552 [hep-ph]} \BibitemShut
  {NoStop}%
\bibitem [{\citenamefont {Devlani}\ \emph {et~al.}(2014)\citenamefont
  {Devlani}, \citenamefont {Kher},\ and\ \citenamefont
  {Rai}}]{Devlani:2014nda}%
  \BibitemOpen
  \bibfield  {author} {\bibinfo {author} {\bibfnamefont {N.}~\bibnamefont
  {Devlani}}, \bibinfo {author} {\bibfnamefont {V.}~\bibnamefont {Kher}},\ and\
  \bibinfo {author} {\bibfnamefont {A.~K.}\ \bibnamefont {Rai}},\ }\href
  {https://doi.org/10.1140/epja/i2014-14154-2} {\bibfield  {journal} {\bibinfo
  {journal} {Eur. Phys. J. A}\ }\textbf {\bibinfo {volume} {50}},\ \bibinfo
  {pages} {154} (\bibinfo {year} {2014})}\BibitemShut {NoStop}%
\end{thebibliography}%

\end{document}